\documentstyle[12pt,nath]{article}

\input{epsf.tex}
\input{rotate}

\newsavebox{\texta}
\newsavebox{\textb}
\newsavebox{\textc}
\newsavebox{\textd}

\newcommand{\saywhere}{
\vspace*{-100pt}
\begin{center}
{\scriptsize Prepared in the fall of 1994 for the proceedings of the\\
19th International Nathiagali Summer College on Physics and Contemporary
Needs, Nathiagali 1994\\
edited by S. A. Ahmad and S. M. Farooqi\\[-5pt] 
(should have been published by Pak Book Company, 
but did never appear in print)}
\end{center}
\vspace{20pt}
}

\begin{document}

\saywhere

\renewcommand{\thefootnote}{\alph{footnote}}

\begin{center}
{\large\bf ELEMENTS OF MICROMASER PHYSICS}\\[2\baselineskip]
Berthold--Georg Englert\\[\baselineskip]
{\em Sektion Physik, Universit\"at M\"unchen\\ Am Coulombwall 1,
 D-85748 Garching\/}\footnote{Also at Max-Planck-Institut f\"ur Quantenoptik,
Hans-Kopfermann-Stra\ss e 1, D-85748 Garching.}
\end{center}

\vfill

\begin{quote}\small \baselineskip=12pt
\begin{center} Abstract \end{center}
The elements of micromaser physics are reviewed in a tutorial way. The
emphasis is on the basic theoretical concepts, not on technical details or
experimental subtleties. After a brief treatment of the atom-photon
interaction according to the Jaynes-Cummings model, the master equation that
governs the dynamics of the one-atom maser is derived. Then the more
important properties of the steady states of the one-atom maser are
discussed, including the trapped states.  The approximations, upon which the
standard theoretical model of the one-atom maser is based, are exhibited. The
methods by which one calculates statistical properties of the emerging atoms
are hinted at.
\end{quote}

\renewcommand{\thefootnote}{\arabic{footnote}}
\setcounter{footnote}{0}

\vfill

\noindent%
This is a tutorial review of micromaser physics, mainly from the point of
view of a theoretician. The audience aimed at consists of non-experts who
want to get an idea about this field of research without perhaps having the
time to go into the finer details. More technical reviews are, of course,
available in the published literature. I would like to draw special
attention to the recent articles of the Garching group~\cite{rev:MPQ} and of
the Paris group~\cite{rev:ENS}, which are both contained in an up-to-date
collection of essays on cavity quantum electrodynamics~\cite{CQED}. In these
articles, the reader will find very many references to papers on various
subtle aspects of micromaser physics that are well beyond the scope of this
tutorial review. There are also textbooks on quantum optics~\cite{QOtexts}
that devote some space to a treatment of the micromaser. Some more recent
work that is not included in these reviews or textbooks will be cited in the
text.

\newpage

\SEC{Experimental setup and other introductory remarks}

\begin{figure}[tb]
\begin{center}
\begin{tabular}{c}
\put(32,10){\footnotesize velocity}
\put(32,0){\footnotesize selector}
\put(114,-4){\footnotesize laser}
\put(104,-14){\footnotesize excitation}
\put(160,-3){\footnotesize resonator with}
\put(170,-13){\footnotesize TE mode}
\put(267,73){\footnotesize state selective}
\put(277,63){\footnotesize ionization}
\put(265,6){\footnotesize upper}
\put(270,-4){\footnotesize level}
\put(301,6){\footnotesize lower}
\put(301,-4){\footnotesize level}
\put(295,-6){\line(0,1){20}}
\leavevmode \epsfxsize=5in \epsffile[14 99 524 212]{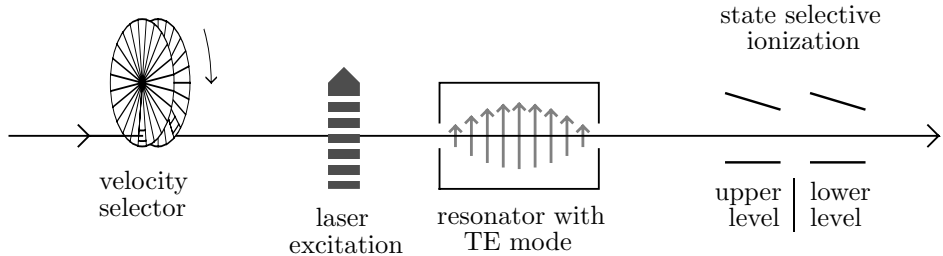}
\end{tabular}
\end{center}
\caption[Experimental setup]{\label{setup}\small\baselineskip=12pt
 Schematic setup of micromaser experiments.}
\end{figure}

The most important parts of the setup of a micromaser experiment are sketched
in Fig.~\ref{setup}. 
A beam of neutral alkaline atoms
(rubidium is the most popular species) is run through a resonator, which has
a very large quality factor to ensure a long photon lifetime. The high
quality is achieved by superconducting surfaces, for which purpose the
resonator is made of niobium and cooled below the critical temperature of
4.2\,K. This low temperature has the additional benefit that the number of
thermal photons is very small (remember the Planck radiation) and thus the
thermal noise is reduced. 

The resonator is tuned such that it possesses a mode in the microwave regime
(of TE type in Fig.~\ref{setup}) which is resonant with an atomic transition
between two Rydberg levels. It is this privileged mode that we shall be
interested in exclusively. The frequencies of all other resonator modes are
very far away on the scale set by the linewidth of the privileged mode,
which is extremely narrow owing to the high resonator quality. 

We refer to said atomic transition between the two distinguished Rydberg
states as the masing transition. Prior to entering the resonator the atoms
are excited to (one of) these Rydberg states. Transitions between Rydberg
states with principal quantum numbers in the 50s to 70s are in the
convenient microwave range, where resonators of ultrahigh quality are
available. In addition, Rydberg atoms are very large (diameters of a few
thousand \AA ngstr\"om units are typical) and therefore possess enormous
dipole moments, so that a strong coupling of the atom to the photons in the
privileged mode is achieved. It is then possible that the atoms emit
microwave photons with a large probability. Since the photons are stored for
a rather long time in the resonator --- lifetimes of a considerable fraction
of a second have been realized experimentally --- even a faint beam of atoms
can pump the resonator effectively. In this way, a maser is operated in which
single atoms traversing the resonator provide for an efficient pump. One is
then dealing with a microscopic maser indeed, a {\em micromaser\/}.

The excitation of the atoms to the Rydberg states of interest is typically
achieved by a laser beam which is crossed by the atoms cross just before they
enter the resonator. It is advantageous to select a narrow velocity group
in the atom beam because then all atoms interact for the same well-controlled
time with the resonator photons. A velocity selection can be performed with
the aid of a set of Fizeau wheels, as indicated in Fig.~\ref{setup}, or by
other means (see the more technical review articles).

The high quality of the resonator is quite essential for all micromaser
experiments. Therefore one cannot perform any direct measurements on the
stored photon field. The only information available is the state of the
emerging atoms. Ideally they can solely end up in either one of the states of
the masing transition. The atoms are probed by state selective ionization in
inhomogeneous electric fields, symbolically indicated in Fig.~\ref{setup} by
two pairs of nonparallel condensor plates. At the first stage only atoms in
the upper masing level are ionized, in the second stage only those in the
lower level. The stripped-off electrons are detected and thus the final state
of the atom is determined.

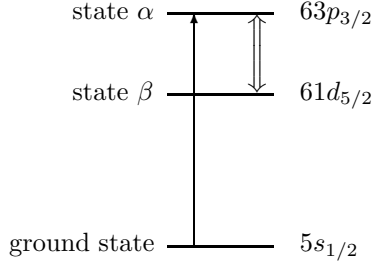
\begin{figure}[tb]
\begin{picture}(390,110)(-165,-30)
\put(50,50){$\Bigg\Updownarrow$}
\thinlines
\put(30,-20){\vector(0,1){88.5}}
\thicklines
\put(20,37.5){\line(1,0){40}}
\put(20,68){\line(1,0){40}}
\put(20,-20){\line(1,0){40}}
\put(-40,-22){\footnotesize ground state}
\put(70,-22){\footnotesize $5s_{1/2}$}
\put(-16,35.5){\footnotesize state $\beta$}
\put(70,35.5){\footnotesize $61d_{5/2}$}
\put(-16,66){\footnotesize state $\alpha$}
\put(70,66){\footnotesize $63p_{3/2}$}
\end{picture}
\caption[Relevant Rb levels]{\label{Rb85}\small\baselineskip=12pt
Levels of $^{85}$Rb relevant for some of the Garching micromaser
experiments. The laser excitation from the ground state to the upper masing
level $\alpha$ is indicated by the $\uparrow$ arrow, the masing transition
by the $\Updownarrow$ arrow. The level spacing is not drawn to scale.
\hrulefill}
\end{figure}

The levels of the $^{85}$Rb isotope that are most relevant for many of the
micromaser experiments in Garching are depicted in Fig.~\ref{Rb85}.
As indicated, the masing levels will be called $\alpha$ and
$\beta$ throughout. In this figure, the excitation from the ground state to
the state $\alpha$ is shown as a one-step process. That can be done by a
suitably tuned, frequency doubled, ring dye laser. But different excitation
schemes, that involve intermediate levels, have also been used successfully.

The heart of the experiment is the interaction of the excited atom with the
resonator photons. To a reasonably good approximation only the two Rydberg
levels of the masing transition are significant for the dynamics during the
passage of the atom through the resonator. We shall, therefore, simplify
matters and think of the masing atoms as two-level systems. The state vectors
$\ket{\UP}\equiv\ket{\alpha}$ and $\ket{\DN}\equiv\ket{\beta}$,
respectively, are then used to denote the excited and the deexcited state of
such a two-level atom.

\SEC{Atom-photon interaction (Jaynes-Cummings model)}

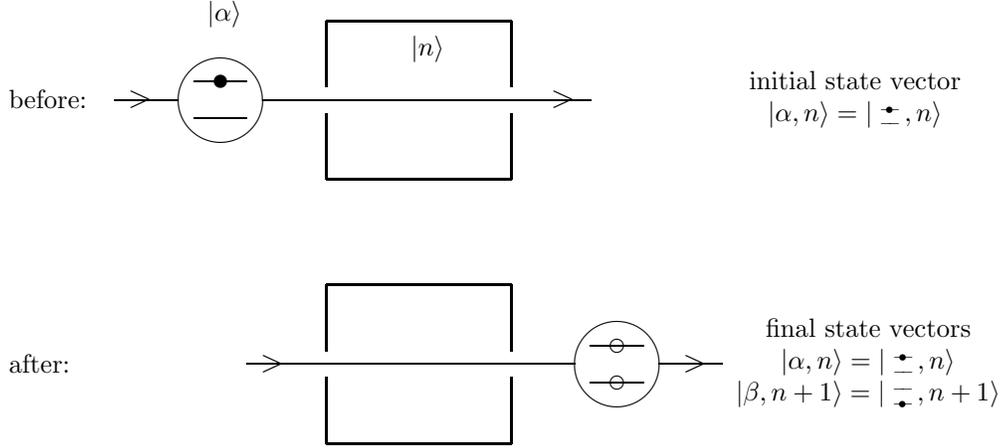
\begin{figure}[tb]
\begin{picture}(390,200)(-40,0)
\put(-30,147){\footnotesize before:}
\put(45,180){\footnotesize $|\alpha\rangle$}
\put(122,167){\footnotesize $|n\rangle$}
\put(250,147)
{\parbox[c]{80pt}{\footnotesize 
\begin{center}initial state vector\\
 $|\alpha,n\rangle=|\up,n\rangle$\end{center}}}
\put(50,150){\circle{30}}
\put(40,143){\line(1,0){20}}
\put(40,157){\line(1,0){20}}
\put(50,157){\circle*{5}}
\put(10,150){\line(1,0){24}}
\put(66,150){\line(1,0){124}}
\put(15,147.2){$>$}
\put(175,147.2){$>$}
\thicklines
\put(90,155){\line(0,1){25}}
\put(160,155){\line(0,1){25}}
\put(90,180){\line(1,0){70}}
\put(90,145){\line(0,-1){25}}
\put(160,145){\line(0,-1){25}}
\put(90,120){\line(1,0){70}}
\put(-30,47){\footnotesize after:}
\put(245,47)
{\parbox[c]{100pt}{\footnotesize 
\begin{center}final state vectors\\
 $|\alpha,n\rangle=|\up,n\rangle$\\$|\beta,n+1\rangle=|\dn,n+1\rangle$
\end{center}}}
\thinlines
\put(200,50){\circle{30}}
\put(190,43){\line(1,0){20}}
\put(190,57){\line(1,0){20}}
\put(200,57){\circle{5}}
\put(200,43){\circle{5}}
\put(216,50){\line(1,0){24}}
\put(60,50){\line(1,0){124}}
\put(65,47.1){$>$}
\put(225,47.1){$>$}
\thicklines
\put(90,55){\line(0,1){25}}
\put(160,55){\line(0,1){25}}
\put(90,80){\line(1,0){70}}
\put(90,45){\line(0,-1){25}}
\put(160,45){\line(0,-1){25}}
\put(90,20){\line(1,0){70}}
\end{picture}
\caption[Before/after]{\label{passage}\small\baselineskip=12pt
Before and after the passage of a two-level atom that is initially excited
through a resonator with $n$ photons initially. \hrulefill}
\end{figure}

A symbolic representation of the passage of a two-level atom through the
resonator is shown in Fig.~\ref{passage}. 
Initially the atom
is in state $\alpha$ and there are supposedly $n$ photons in the resonator,
so that the initial state vector of the combined atom-field system is
$\ket{\alpha,n}=\ket{\UP,n}$. After the interaction we may have either one of
two final state vectors --- $\ket{\alpha,n}$ if no photon has been emitted,
or $\ket{\beta,n+1}=\ket{\DN,n+1}$ if a photon has been emitted --- or a
superposition of both.

For a calculation of the actual final state we use the standard {\em
Jaynes-Cummings model\/}\footnote{The name derives from a 1963
paper~\cite{JC63}.} for the theoretical description of the atom-field
dynamics. The fundamental variables of the privileged field mode are the
photon ladder operators $a$ and $a^{\dag}$ with their well known properties,
of which
\begin{eqns}{rcl}
[a,a^{\dag}]=1 \,,\,\N\ket{n}&=&\ket{n}n\\
a^{\dag}\ket{n}&=&\ket{n+1}\sqrt{n+1}
\label{phlad}
\end{eqns}%
are perhaps the most important ones. Likewise we employ ladder operators
$\sigma$ and $\sigma^{\dag}$ for the two-level atoms. Their defining
properties are
\begin{eqns}{rclcrcl}
\sigma\ket{\alpha}&=&\ket{\beta}&\mbox{\ or\ }
&\sigma\ket{\UP}&=&\ket{\DN}\,,\\
\sigma^{\dag}\ket{\beta}&=&\ket{\alpha}&\mbox{\ or\ }
&\sigma^{\dag}\ket{\DN}&=&\ket{\UP}\,,
\end{eqns}%
and
\begin{eqns}{rclcrcl}
\sigma\ket{\beta}&=&0&\mbox{\ or\ } &\sigma\ket{\DN}&=&0\,,\\
\sigma^{\dag}\ket{\alpha}&=&0&\mbox{\ or\ } &\sigma^{\dag}\ket{\UP}&=&0\,,
\end{eqns}%
with the consequences
\begin{equation}
\sigma^2=0\,,\ \sigma^{\dag2}=0
\end{equation}
as well as
\begin{eqns}{rcccl}
\sigma^{\dag}\sigma&=&\ket{\alpha}\bra{\alpha}&=&\ket{\UP}\bra{\UP}\,,\\
\sigma\sigma^{\dag}&=&\ket{\beta}\bra{\beta}&=&\ket{\DN}\bra{\DN}\,.
\label{atlad}
\end{eqns}%
The latter are useful ways of expressing the projectors to the atomic states.

\begin{figure}[tb]
\begin{picture}(390,270)(0,0)
\put(50,50){\epsffile[100 60 400 250]{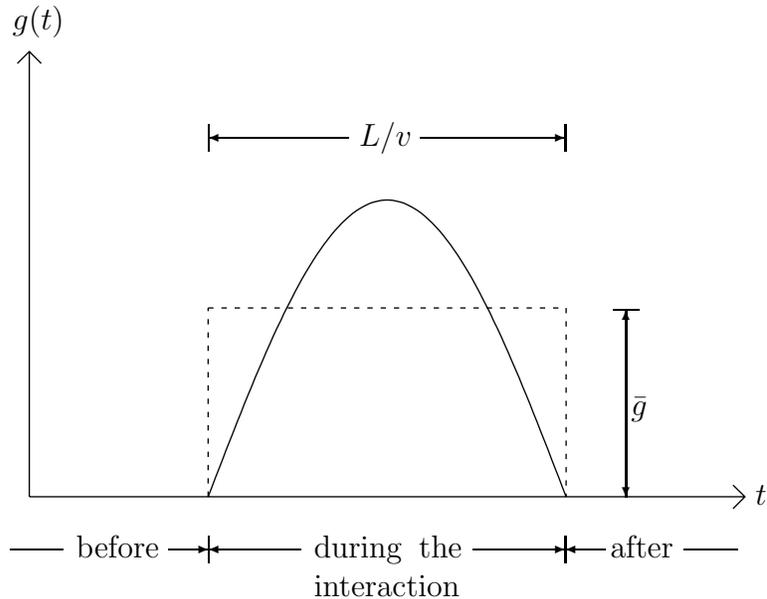}}
\put(342,58){$t$}
\put(61,238){$g(t)$}
\put(135,192){\line(0,1){10}}
\put(270,192){\line(0,1){10}}
\put(192,194){$L/v$}
\put(188,197){\vector(-1,0){53.2}}
\put(215,197){\vector(1,0){55.2}}
\put(288,132){\line(1,0){10}}
\put(295,92){$\bar{g}$}
\put(293,100){\vector(0,1){32.2}}
\put(293,100){\vector(0,-1){39.2}}
\put(135,36){\line(0,1){10}}
\put(270,36){\line(0,1){10}}
\put(175,38){\parbox[t]{56pt}{during the interaction}}
\put(170,41){\vector(-1,0){35.2}}
\put(235,41){\vector(1,0){35.2}}
\put(85,38){\parbox[t]{32pt}{before}}
\put(120,41){\vector(1,0){15.2}}
\put(80,41){\line(-1,0){20}}
\put(287,38){\parbox[t]{30pt}{after}}
\put(285,41){\vector(-1,0){15.2}}
\put(315,41){\line(1,0){20}}
\end{picture}
\caption[Rabi freqency as a function of time]{\label{goft}
\small\baselineskip=12pt
Time dependent Rabi frequency $g(t)$ (solid line) and effective Rabi
frequency $\bar{g}$. \hrulefill}
\end{figure}

In the Jaynes-Cummings model the dynamics is generated by the Hamilton
operator
\savebox{\texta}{photon}
\savebox{\textb}{atom}
\savebox{\textc}{interaction}
\begin{equation}\label{JCMham}
H=\underbrace{\hbar\omega\N\rule[-6pt]{0pt}{5pt}}_{\usebox{\texta}}
+\underbrace{\hbar\Omega\sigma^{\dag}\sigma
\rule[-6pt]{0pt}{5pt}}_{\usebox{\textb}}
-\underbrace{\hbar g(t)(\G)\rule[-6pt]{0pt}{5pt}}_{\usebox{\textc}}\,.
\end{equation}
Here, $\hbar\omega$ is the energy per photon, $\hbar\Omega$ is the atomic
level spacing (the deexcited atom is assigned energy zero by convenient
convention), and the so-called Rabi frequency $g(t)$ measures the strength of
the interaction. Inasmuch as the atom enters and leaves the resonator, this
Rabi frequency possesses a natural time dependence. It vanishes as long as
the atom is outside the resonator, and is large while the atom is inside. To
the extent to which the atomic center-of-mass motion is
classical,\footnote{This condition is very well obeyed under the ordinary
experimental conditions. Under extreme circumstances, however, the quantum
nature of the center-of-mass motion may become important~\cite{slowatoms}.}
the duration of the interaction is given by the ratio $L/v$ of the cavity
length $L$ and the atomic velocity $v$. These circumstances are sketched in
Fig.~\ref{goft} 
where we also indicate the effective Rabi
frequency $\bar{g}$, which is given by the time average of $g(t)$,
\begin{equation}\label{gbar}
\bar{g}=\frac{v}{L}\int\!\diff{t}\,g(t)\,.
\end{equation}
To be specific let us report typical numbers for the micromaser experiments
in Garching that make use of the Rydberg transition of Fig.~\ref{Rb85}. The
resonator is of cylindrical shape with an almost circular cross section of
diameter $\sim$24\,mm and a length $L$ that is also $\sim$24\,mm. The
experiments use the photons in the TE$_{121}$ mode. Here one has the
frequency values
\begin{eqns}{rcl}\label{freqnumb}
\omega\cong\Omega&\cong&2\pi\times21.5\,\mbox{GHz}\,,\\
\bar{g}&\cong&44\,\mbox{kHz}=\pi\times14\,\mbox{kHz}\,,
\end{eqns}%
which correspond to the energy values
\begin{eqns}{rcl}\label{enernumb}
\hbar\omega\cong\hbar\Omega&\cong&9\times10^{-5}\,\mbox{eV}\,,\\
\hbar\bar{g}&\cong&3\times10^{-11}\,\mbox{eV}\,.
\end{eqns}%
Please note how small is the energy per photon\footnote{Recall that visible
photons have energies of 1--2\,eV.} and how tiny is the interaction energy.

The transition operator 
\begin{equation}
\gamma\equiv\G
\end{equation}
effects atomic transitions along with the emission or absorption of a photon:
\begin{eqns}{rcl}
\gamma\ket{\UP,n}&=&\ket{\DN,n+1}\sqrt{n+1}\,,\\
\gamma\ket{\DN,n+1}&=&\ket{\UP,n}\sqrt{n+1}\,.
\end{eqns}%
The eigenvalues of $\gamma$ are 
$\gamma'=0,\pm1\pm\sqrt{2},\pm\sqrt{3}\ldots$, and its eigenstates
(so-called `dressed states' of one kind in the jargon of quantum optics) are
very simple linear combination of the atom-field states that we have been
using so far, viz.
\begin{eqns}{rclcrcl}
\ket{\gamma_0}&\equiv&\ket{\DN,0}\,,&&\ket{\gamma^{\pm}_n}&\equiv&
\frac{1}{\sqrt{2}}\left(\ket{\UP,n}\pm\ket{\DN,n+1}\right)\,,\\
\gamma\ket{\gamma_0}&=&0\,,&&
\gamma\ket{\gamma^{\pm}_n}&=&\ket{\gamma^{\pm}_n}\left(\pm\sqrt{n+1}\right)
\,.
\end{eqns}%
They are mutually orthogonal and normalized to unity,
\begin{equation}\renewcommand{\arraystretch}{1.4}
\braket{\gamma'}{\gamma''}=\delta(\gamma',\gamma'')
=\left\{
\begin{array}{rcccl}
1&&\mbox{if}&& \gamma'=\gamma''\\
0&&\mbox{if}&& \gamma'\neq\gamma''
\end{array}
\right.
\end{equation}
and complete,
\begin{equation}
\sum_{\gamma'}\ket{\gamma'}\bra{\gamma'}=1\,,
\end{equation}
and constitute a convenient basis for expanding arbitrary state vectors.

The important identity
\begin{equation}
\gamma^2=\N+\sigma^{\dag}\sigma
\end{equation}
is easily demonstrated with the aid of the algebraic properties of the ladder
operators stated in Eqs.~(\ref{phlad})--(\ref{atlad}). We use it to rewrite
the Hamilton operator (\ref{JCMham}) into a form,
\begin{equation}
H=\hbar\omega\gamma^2+\hbar(\Omega-\omega)\sigma^{\dag}\sigma
-\hbar g(t)\gamma\,,
\end{equation}
which is particularly transparent on resonance ($\omega=\Omega)$:
\begin{equation}
H(t)=\hbar\omega\gamma^2-\hbar g(t)\gamma\,.
\end{equation}
For the sake of simplicity, we shall confine ourselves to this resonant
situation for the rest of this tutorial review.

Since the transition operator is then a constant of motion --- the vanishing
commutator $[H(t),\gamma(t)]=0$ implies $\gamma(t)=\gamma(t_0)$ --- the
Schr\"odinger equation obeyed by the eigenstates of $\gamma(t)$ is very
simple. For the left eigenstates, it reads
\begin{equation}
i\hbar\partt\bra{\gamma',t}
=\bra{\gamma',t}H(t)
=\left[\hbar\omega\gamma'^2-\hbar g(t)\gamma'\right]\bra{\gamma',t}\,.
\end{equation}
and can be solved immediately. The outcome is
\begin{equation}\label{evolution}
\bra{\gamma',t}
=\underbrace{\exp\Bigl[-i\omega\gamma'^2(t-t_0)
\Bigr]\rule[-10pt]{0pt}{10pt}}_{\mbox{free evolution}}
\underbrace{\exp\Bigl[i\gamma'
\int^t_{t_0}\diff{t'}\,g(t')\Bigr]}_{\mbox{interaction}}
\bra{\gamma',t_0}\,.
\end{equation}
We note that the evolution is naturally split into two contributions: the
free evolution and what results from the interaction. 

We are particularly interested in the overall change asked for in the context
of Fig.~\ref{passage}. Then the instants $t_0$ and $t$ are `before' and
`after' the interaction, so that the integral in (\ref{evolution}) equals the
effective Rabi frequency of (\ref{gbar}) times the classical interaction time
$L/v$,
\begin{equation}
\int^t_{t_0}\diff{t'}\,g(t')=\bar{g}L/v\equiv\varphi\,.
\end{equation}
For obvious reasons, the quantity $\varphi$ here introduced is called the
accumulated Rabi angle. This single number summarizes the net effect of the
atom-field interaction.

We have thus found a first central result of micromaser theory:
\begin{equation}\label{JCMchange}
\fbox{\parbox{220pt}{
\begin{center}
\hspace*{10pt}Resonant interaction leads to a change\hspace*{10pt}
\begin{displaymath}
\renewcommand{\arraystretch}{1.4}
\begin{array}{rcl}
\bra{\gamma'}&\to&e^{i\varphi\gamma'}\bra{\gamma'}\\
\ket{\gamma'}&\to&\ket{\gamma'}e^{-i\varphi\gamma'}
\end{array}
\end{displaymath}
on top of the free evolution.
\end{center}
}}
\end{equation}
As an application let us reconsider the situation of Fig.~\ref{passage}.
First, we write the initial state vector $\ket{\UP,n}$ as a superposition of
eigenvectors of the interaction operator $\gamma$,
\begin{eqns}{rcl}
\ket{\UP,n}&=&\frac{1}{2}\left(\ket{\UP,n}+\ket{\DN,n+1}\right)
+\frac{1}{2}\left(\ket{\UP,n}-\ket{\DN,n+1}\right)\\
&=&\frac{1}{\sqrt{2}}\left(\ket{\gamma^+_n}+\ket{\gamma^-_n}\right)\,.
\end{eqns}%
Second, we make use of (\ref{JCMchange}) to find the change resulting from
the interaction,
\begin{equation}
\ket{\UP,n}\INT
\frac{1}{\sqrt{2}}\left[\ket{\gamma^+_n}\exp(-i\varphi\sqrt{n+1})
+\ket{\gamma^-_n}\exp(i\varphi\sqrt{n+1})\right]\,.
\end{equation}
Third, we express the outcome in terms of the atom-field states,
\begin{equation}\label{OAchange}
\ket{\UP,n}\INT
\ket{\UP,n}\cos(\varphi\sqrt{n+1})
+\ket{\DN,n+1}[-i\sin(\varphi\sqrt{n+1})]\,,
\end{equation}
which is the desired result. In particular, it states the probabilities with
which the two possible final atom-field states occur,
\begin{equation}\label{transprob}
\renewcommand{\arraystretch}{1.4}
\ket{\UP,n}\INT\left\{
\begin{array}{ccl}
\ket{\UP,n}&\mbox{with probability}&\cos^2(\varphi\sqrt{n+1})\,,\\
\ket{\DN,n+1}&\mbox{with probability}&\sin^2(\varphi\sqrt{n+1})\,.
\end{array}
\right.
\end{equation}

It is essential to realize that this result is dramatically different from
the one in the usual spontaneous-and-induced emission process. There the
physics is well described by the lowest-order perturbation theory, in which
the argument $\varphi\sqrt{n+1}$ of the sine and cosine function in
(\ref{transprob}) is infinitesimal and individual transitions are not
happening frequently. In addition, usual spontaneous emission involves not
only just one mode of the radiation field but a continuous range of modes
with neighboring frequencies. In summary, the photon emission in a
micromaser experiment has nothing in common with the much more familiar
spontaneous emission by atoms in free space.

We note that there are special situations in which the photon emission
happens with certainty,
\begin{eqns}{rcl}
\ket{\UP,n}\INT\ket{\DN,n+1}&\mbox{if}&\varphi\sqrt{n+1}=\pi/2\\
&&(\mbox{e.\,g.}\ n=0\,,\ \varphi=\pi/2)\,,
\end{eqns}%
or in which surely no emission takes place,
\begin{eqns}{rcl}\label{emitnot}
\ket{\UP,n}\INT\ket{\UP,n}&\mbox{if}&\varphi\sqrt{n+1}=\pi\\
&&(\mbox{e.\,g.}\ n=0\,,\ \varphi=\pi)\,.
\end{eqns}%
Of course, one could add integer multiples to $\varphi\sqrt{n+1}$ in both
cases without changing the final states. Clearly, these special situations
are unknown in the context of the usual spontaneous emission. In the
micromaser the atom can first emit and then reabsorb the photon --- this is
the case of (\ref{emitnot}) --- thus undergoing a complete Rabi cycle, or
perhaps a few complete cycles.

\SEC{One-atom operation}

The beam of atoms in Fig.~\ref{setup} comes originally from an opening in an
oven, has been collimated, and perhaps velocity selected, before it is
directed through the resonator. If no special action is taken, as is the
standard experimental situation, the atoms in the beam are uncorrelated. In
other words, the atoms arrive at random. The probability $p(t)\diff{t}$ that
two successive atoms are separated in time by $t\cdots t+\diff{t}$ is thus
given by the Poisson formula\footnote{The Poisson formula governs also more
mudane phenomena, such as the temporal spacing between drops of rain.}
\begin{equation}\label{waiting}
p(t)\diff{t}=e^{-rt}r\diff{t}\,,
\end{equation}
where $r$ is the rate at which the atoms arrive. The average temporal spacing
between atoms is clearly equal to $1/r$; the average spatial distance is
then $v/r$ if all atoms move at the same speed $v$, a condition that we
shall assume to be obeyed for the time being.

\begin{figure}[tb]
\begin{picture}(390,120)(-65,-15)
\thicklines
\put(90,55){\line(0,1){25}}
\put(160,55){\line(0,1){25}}
\put(90,80){\line(1,0){70}}
\put(90,45){\line(0,-1){25}}
\put(160,45){\line(0,-1){25}}
\put(90,20){\line(1,0){70}}
\thinlines
\put(90,82){\line(0,1){10}}
\put(160,82){\line(0,1){10}}
\put(122,84){\footnotesize $L$}
\put(120,87){\vector(-1,0){30.2}}
\put(130,87){\vector(1,0){30.2}}
\put(-40,50){\line(1,0){330}}
\put(-30,47.1){$>$}
\put(277,47.1){$>$}
\put(120,50){\circle*{7}}
\put(117,54){$\to$}
\put(118,60){\footnotesize $v$}
\put(0,50){\circle*{7}}
\put(-3,54){$\to$}
\put(-2,60){\footnotesize $v$}
\put(220,50){\circle*{7}}
\put(217,54){$\to$}
\put(218,60){\footnotesize $v$}
\put(250,50){\circle*{7}}
\put(247,54){$\to$}
\put(248,60){\footnotesize $v$}
\put(121,15){$\underbrace{\hspace*{98pt}}_{\mbox{\footnotesize$t_+>L/v$}}$}
\put(1,15){$\underbrace{\hspace*{118pt}}_{\mbox{\footnotesize$t_->L/v$}}$}
\end{picture}
\caption[One-atom event]{\label{OAevent}\small\baselineskip=12pt
An atom gives rise to a one-atom event if the temporal separations from its
predessor, $t_+$, and from its successor, $t_-$, both exceed the classical
interaction time $L/v$, equal to the ratio of the cavity length $L$ and the
atomic velocity $v$. \hrulefill}
\end{figure}
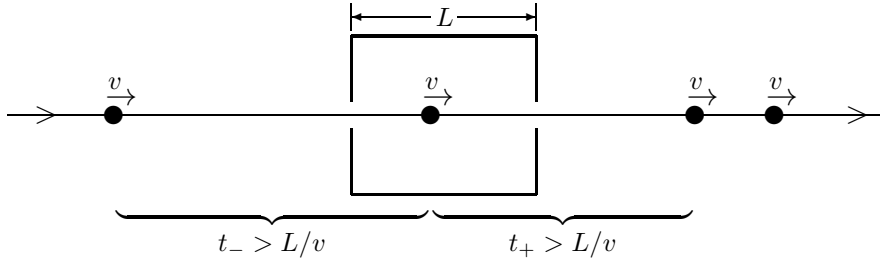

If the beam is faint enough, most of the atoms will enter the resonator after
the preceding atom has left it, and will leave before the succeeding atom
arrives. This is the situation depicted in Fig.~\ref{OAevent}.
The probability that the middle atom in this figure gives rise to such a
one-atom event is equal to the product of the probabilities that both
spacings, $t_-$ and $t_+$, exceed the classical interaction time $L/v$. Now,
the Poisson statistics (\ref{waiting}) imply that the probability for a
spacing greater than a certain specified period $T$ is given by
\begin{equation}
\int_T^{\infty}\diff{t}\,p(t)=e^{-rT}\,.
\end{equation}
Consequently, the asked-for probability of a one-atom event is
\begin{equation}\label{OAprob}
\underbrace{\exp(-rL/v)}_{t_->L/v}\ \times\
\underbrace{\exp(-rL/v)}_{t_+>L/v}\ =\ \exp(-2rL/v)\,.
\end{equation}
A typical number for $L/v$ is 50\,$\mu$s, so that at the rather large beam
rate of $r\cong1000/$s this probability is about $\exp(-0.1)\cong90\%$. In
other words, 90\% of the atoms would contribute to one-atom events. More
typical is a beam rate of $r\cong10/$s, for which the probability for
one-atom events is about $\exp(-0.001)=99.9\%$. Collective events, in which
more than one atom interact with the resonator photons at the same time, are
then extremely rare and can be justifiably disregarded. We are then dealing
with the {\em One-Atom Maser\/}, the most intriguing version of the
micromaser.

\SEC{One-atom maser: Dynamics}

For the theoretical treatment of the dynamics of the photon field in the
one-atom maser (OAM) we employ the standard master equation approach. It
deals with a coarse grain time evolution of the photon field. Changes in the
field that occur on the short time scale, on which the atom-field interaction
happens, are ignored. In essence, one is only interested in the changes that
occur in the photon state during the passage of a fair number of atoms. 

Since we are focussing on the state of the photon field, rather than on the
state of the entire system consisting of very many atoms and the photon
field, we are treating a partial system. More precisely: we are treating an
open, driven quantum system. Its state cannot be described in terms of a
state vector $\ket{\ }$ or (one of) its wavefunctions. What one needs is
a state operator $\rho$, which is a positive operator of unit trace. 
These two properties of $\rho$ ensure that all probabilities 
$\bra{\ }\rho\ket{\ }$ are non-negative and that the sum of all probabilities
of a complete set of mutually exclusive alternatives is unity.

The matrix elements of $\rho$ make up the so-called
density matrix; for instance, when the photon number states $\bra{n}$ and
$\ket{m}$ are used, we get the number-state density matrix
$\rho_{nm}=\bra{n}\rho\ket{m}$. Just like there are very many wavefunctions
to one state vector $\ket{\ }$, there are also very many density matrices
to one state operator $\rho$. It is clearly advantageous to concentrate on
the (somewhat abstract) state operator in the first place, and to choose a
convenient matrix representation later when it is really needed.

For the photon field inside the resonator of a OAM, the state operator
$\rho$ --- we shall frequently speak of `the state $\rho$' simply --- is a
time-dependent function of the ladder operators $a^{\dag}$ and $a$:
$\rho=\rho_t(a^{\dag},a)$. For the benefit of more transparent equations, we
shall suppress the arguments of $\rho$ where misunderstandings are unlikely.
The master equation obeyed by $\rho$ has the general structure
\begin{equation}\label{OAMmaster}
\partt\rho=
\partt\rho_{\Big\vert_{\stext{free}}}
+\partt\rho_{\Big\vert_{\stext{gain}}}
+\partt\rho_{\Big\vert_{\stext{loss}}}\,.
\end{equation}
The `free' contribution is, of course, generated by the photon part of the
Hamilton operator (\ref{JCMham}),
\begin{equation}\label{MEfree}
\partt\rho_{\Big\vert_{\stext{free}}}
=\frac{1}{i\hbar}\left[H_{\stext{photon}},\rho\right]
=-i\omega[\N,\rho]\,.
\end{equation}
The `gain' contribution to (\ref{OAMmaster}) is the additional change
resulting from the interacting atoms traversing the resonator, and the `loss'
contribution originates in the coupling of the photons to their surroundings.
We shall first derive the gain term and then turn to the loss term.

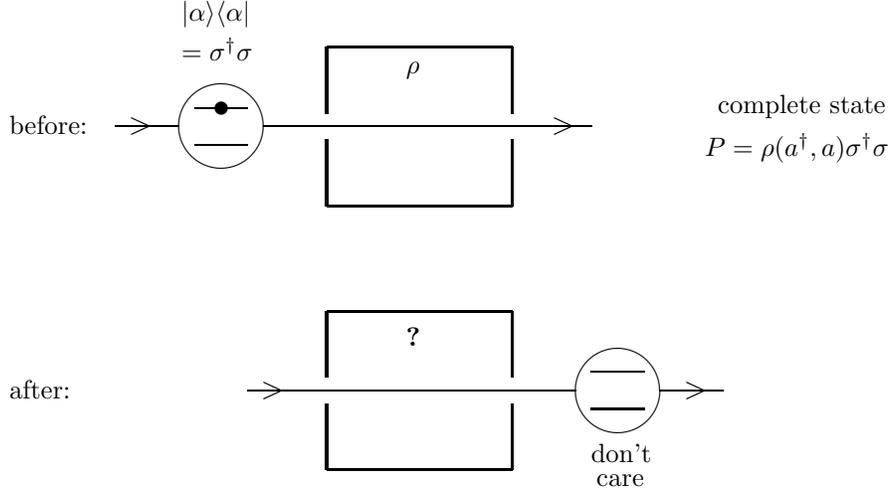
\begin{figure}[tb]
\begin{picture}(390,210)(-55,0)
\put(-30,147){\footnotesize before:}
\put(35,190){\footnotesize $|\alpha\rangle\langle\alpha|$}
\put(35,175){\footnotesize $=\sigma^{\dag}\sigma$}
\put(120,170){\footnotesize $\rho$}
\put(238,155){\footnotesize complete state} \put(233,139){\footnotesize
$P=\rho(a^{\dag},a)\sigma^{\dag}\sigma$}
\put(50,150){\circle{30}}
\put(40,143){\line(1,0){20}}
\put(40,157){\line(1,0){20}}
\put(50,157){\circle*{5}}
\put(10,150){\line(1,0){24}}
\put(66,150){\line(1,0){124}}
\put(15,147.2){$>$}
\put(175,147.2){$>$}
\thicklines
\put(90,155){\line(0,1){25}}
\put(160,155){\line(0,1){25}}
\put(90,180){\line(1,0){70}}
\put(90,145){\line(0,-1){25}}
\put(160,145){\line(0,-1){25}}
\put(90,120){\line(1,0){70}}
\put(-30,47){\footnotesize after:}
\put(120,67){\footnotesize\bf ?}
\put(190,23){\footnotesize don't}
\put(192,14){\footnotesize care}
\thinlines
\put(200,50){\circle{30}}
\put(190,43){\line(1,0){20}}
\put(190,57){\line(1,0){20}}
\put(216,50){\line(1,0){24}}
\put(60,50){\line(1,0){124}}
\put(65,47.1){$>$}
\put(225,47.1){$>$}
\thicklines
\put(90,55){\line(0,1){25}}
\put(160,55){\line(0,1){25}}
\put(90,80){\line(1,0){70}}
\put(90,45){\line(0,-1){25}}
\put(160,45){\line(0,-1){25}}
\put(90,20){\line(1,0){70}}
\end{picture}
\caption[OAM gain]{\label{OAMgain}\small\baselineskip=12pt
Gain in OAM operation. The two-level atom is excited when it enters the
resonator; the initial atomic state is
$|\UP\rangle\langle\UP|=|\alpha\rangle\langle\alpha|=\sigma^{\dag}\sigma$.
Prior to the interaction the photon field is in the state
$\rho(a^{\dag},a)$. Thus the initial state of the complete system is
$P=\rho(a^{\dag},a)\sigma^{\dag}\sigma$. The interaction changes this state.
We do not care about the final atomic state, but want to know the photon
state after the interaction. \hrulefill}
\end{figure}

The gain is determined by considering the situation of Fig.~\ref{OAMgain}.
After employing the number-state matrix of the initial field state $\rho$,
\begin{equation}
\rho(a^{\dag},a)=\sum^{\infty}_{n,m=0}\ket{n}\rho_{nm}\bra{m}\,,
\end{equation}
to write the complete initial state $P$ in the form\footnote{The letter $P$
is a capital $\rho$, not a capital $p$.}
\begin{equation}
P=\sum\limits_{n,m}\ket{\UP,n}\rho_{nm}\bra{\UP,m}\,.
\end{equation}
Now we can make use of Eq.~(\ref{OAchange}) to find the effect of the
interaction,
\begin{nareqns}{rccl}
P&\INT&\sum\limits_{n,m}&
\left[\ket{\UP,n}\cos(\varphi\sqrt{n+1})
-\ket{\DN,n+1}i\sin(\varphi\sqrt{n+1})\right]\rho_{nm}\\
&&&\times\left[\cos(\varphi\sqrt{m+1})\bra{\UP,m}
+i\sin(\varphi\sqrt{m+1})\bra{\DN,m+1}\right]\\
&&\equiv&P_{\stext{after}}\,.
\end{nareqns}%
This is the final state operator of the complete system consisting of one
atom and the photon field (we remember that the `free' evolution comes on
top). We do not care about the atom and therefore we trace over the two-level
degree of freedom (symbolically: $\mbox{tr}_{\stext{atom}}$) to obtain the
final photon state,
\begin{equation}
\rho(a^{\dag},a)\INT
\mbox{tr}_{\stext{atom}}\left\{P_{\stext{after}}\right\}\,.
\end{equation}
In more detail this reads
\begin{nareqns}{rl}
\rho=\sum\limits_{n,m}\ket{n}\rho_{nm}\bra{m}&\\
\INT&\sum\limits_{n,m}
\left[\ket{n}\cos(\varphi\sqrt{n+1})\rho_{nm}
\cos(\varphi\sqrt{m+1})\bra{m}\right.\\
&+\left.\ket{n+1}\sin(\varphi\sqrt{n+1})\rho_{nm}
\sin(\varphi\sqrt{m+1})\bra{m+1}\right]\,.
\end{nareqns}%
Upon recalling that $a^{\dag}(aa^{\dag})^{-1/2}$ is a normalized ladder
operator,
\begin{equation}
a^{\dag}(aa^{\dag})^{-1/2}\ket{n}=\ket{n+1}\,,
\end{equation}
one establishes the identities
\begin{eqns}{rcl}
\ket{n}\cos({\varphi\sqrt{n+1}})&=&
\cos[{\varphi(aa^{\dag})^{1/2}}]\ket{n}\,,\\
\ket{n+1}\sin({\varphi\sqrt{n+1}})&=&
a^{\dag}\frac{\mbox{$\sin[{\varphi(aa^{\dag})^{1/2}}]$}}%
{\mbox{$(aa^{\dag})^{1/2}$}}\ket{n}\,.
\end{eqns}%
These are then used to arrive at a second central result of micromaser
theory:
\begin{equation}\label{OAgain}
\fbox{\parbox{280pt}{
\begin{center}
\hspace*{10pt}Resonant interaction of atoms entering in\hspace*{10pt}\\
\hspace*{10pt}the $\UP$ state leads to a change\hspace*{10pt}
\begin{displaymath}
\renewcommand{\arraystretch}{1.4}\arraycolsep=2pt
\begin{array}{rrcl}
\rho\ \longrightarrow&
\cos[\varphi(aa^{\dag})^{1/2}]&\rho&\cos[\varphi(aa^{\dag})^{1/2}]\\
&+a^{\dag}\frac{\mbox{$\sin[{\varphi(aa^{\dag})^{1/2}}]$}}%
{\mbox{$(aa^{\dag})^{1/2}$}}&\rho&
\frac{\mbox{$\sin[{\varphi(aa^{\dag})^{1/2}}]$}}%
{\mbox{$(aa^{\dag})^{1/2}$}}a
\end{array}
\end{displaymath}
\hspace*{10pt}in the photon state on top of the free evolution.\hspace*{10pt}
\end{center}
}}
\end{equation}
We add the remark that an analogous result with $a^{\dag}$ and $a$
interchanged is found for atoms that enter the resonator in the $\DN$ state.
The derivation is left to the reader as a simple exercise.

The gain term in (\ref{OAMmaster}) is now available. It is given by
\begin{eqns}{rl}\label{MEgain}
\mbox{\Large$\partt$}\rho_{\Big\vert_{\stext{gain}}} = & 
\Big[\mbox{beam rate}\Big]\,\times\,\Big[\mbox{change by one atom}\Big]\\
=&r\Bigg\{\cos[\varphi(aa^{\dag})^{1/2}]\,\rho\,%
\cos[\varphi(aa^{\dag})^{1/2}]\\
&\left.+a^{\dag}\frac{\mbox{$\sin[{\varphi(aa^{\dag})^{1/2}}]$}}%
{\mbox{$(aa^{\dag})^{1/2}$}}\,\rho\,%
\frac{\mbox{$\sin[{\varphi(aa^{\dag})^{1/2}}]$}}%
{\mbox{$(aa^{\dag})^{1/2}$}}a-\rho\right\}\,.
\end{eqns}%
This expression is valid for the coarse grain time scale provided that the
arriving atoms are uncorrelated, as we have assumed in Eq.~(\ref{waiting}),
for example.

We turn to the loss term. Here we have to model the coupling to a thermal
reservoir. In view of what we have at hand, the simplest model is that of a
thermal beam of very many, very weakly interacting atoms. This beam is
supposed to be at a certain temperature, so that the rate $r_{\alpha}$ of
$\UP$ atoms is larger than the rate $r_{\beta}$ of the $\DN$ atoms. The
ratio of these partial rates is, of course, given by the Boltzmann factor,
\begin{equation}\label{nudef}
r_{\alpha}/r_{\beta}=\exp\left(-\frac{\hbar\omega}{k_{\stext{B}}T}\right)
\equiv\frac{\nu}{\nu+1}\,,
\end{equation}
where $k_{\stext{B}}$ is the Boltzmann constant and $T$ the temperature.
Anticipating its convenience, we introduce the parametrization in terms of
the positive number $\nu$ whose physical significance will become clear
later.

For the $\UP$ atoms in the beam, we have a contribution to the loss term
which is analogous to the right-hand side of Eq.~(\ref{MEgain}), except that
we are now dealing with a tiny Rabi angle $\phi$ rather than with the
macroscopic angle $\varphi$. The lowest non-vanishing order in $\phi$ is all
that we are going to take into account. Thus we find
\begin{equation}
\partt\rho_{\Big\vert_{\stext{loss,$\alpha$}}}=
r_{\alpha}\phi^2\left[a^{\dag}\rho a -\frac{1}{2}aa^{\dag}\rho
-\frac{1}{2}\rho aa^{\dag}\right]
\end{equation}
for the $\UP$ contribution. The $\DN$ contribution is immediately available
too, because the remark after (\ref{OAgain}) implies that we have to
interchange the ladder operators $a$ and $a^{\dag}$,
\begin{equation}
\partt\rho_{\Big\vert_{\stext{loss,$\beta$}}}=
r_{\beta}\phi^2\left[a\rho a^{\dag} -\frac{1}{2}\N\rho 
-\frac{1}{2}\rho\N\right]\,.
\end{equation}
Upon introducing the loss rate $A\equiv(r_{\beta}-r_{\alpha})\phi^2$ and
the parameter $\nu$ of (\ref{nudef}) we combine these partial answers into
\begin{eqns}{rrl}\label{MEloss}
\mbox{\Large$\partt$}\rho_{\Big\vert_{\stext{loss}}}=&
-\frac{1}{2}A(\nu+1)\left[\N\rho-2a\rho a^{\dag}+\rho\N\right]&\\
&-\frac{1}{2}A\nu\left[aa^{\dag}\rho -2a^{\dag}\rho a 
+\rho aa^{\dag}\right]&\,.
\end{eqns}%
The three ingredients of the master equation (\ref{OAMmaster}) are now at
hand.

The physical significance of the rate $A$ and the number $\nu$ is revealed by
a quick look at the dynamics of an unpumped resonator, for which
\begin{nareqns}{rr}\label{unpumped}
\mbox{\Large$\partt$}\rho_{\Big\vert_{\stext{loss}}}=-i\omega[\N,\rho]&
-\frac{1}{2}A(\nu+1)\left[\N\rho-2a\rho a^{\dag}+\rho\N\right]\\
&-\frac{1}{2}A\nu\left[aa^{\dag}\rho -2a^{\dag}\rho a 
+\rho aa^{\dag}\right]
\end{nareqns}%
is the equation of motion. This master equation is in fact so simple that the
general solution for an arbitrary initial state can be written down in a
number of ways. In the present context that is not so essential. We are
content with these observations:
\begin{list}{(\roman{abc})}{\usecounter{abc}\setlength{\topsep}{0pt}}
\item The stationary state of (\ref{unpumped}) is the thermal photon state
\begin{equation}\label{thermal}
\rho_{\stext{th}}
=\frac{1}{\nu+1}\left(\frac{\nu}{\nu+1}\right)^{\mbox{$\N$}}
=\left[1-\exp\left(-\frac{\hbar\omega}{k_{\stext{B}}T}\right)\right]
\exp\left(-\frac{\hbar\omega}{k_{\stext{B}}T}\N\right)\,,
\end{equation}
which identifies $\nu$ as the number of thermal photons inside the resonator
at temperature $T$.
\item For the evolution of the mean photon field 
$\langle a\rangle=\mbox{tr}\{a\rho\}$ (where
$\mbox{tr}\equiv\mbox{tr}_{\stext{photon}}$ for short) we have
\begin{equation}
\frac{\diff{}}{\diff{t}}\langle a\rangle=\mbox{tr}\left\{a\partt\rho\right\}
=(-i\omega-A/2)\langle a\rangle\,,
\end{equation}
so that
\begin{equation}
\langle a\rangle_t=\langle a\rangle_0e^{-i\omega t}e^{-At/2}\,.
\end{equation}
This shows that the field decays with the rate $A/2$ towards its stationary
null value.
\item Likewise we find for the mean number of photons,
\begin{equation}\label{numbdecay}
\frac{\diff{}}{\diff{t}}\langle\N\rangle
=\mbox{tr}\left\{\N\partt\rho\right\}
=-A(\langle\N\rangle-\nu)\,,
\end{equation}
and therefore
\begin{equation}
\langle\N\rangle_t=\nu+(\langle\N\rangle_0-\nu)e^{-At}\,.
\end{equation}
The mean photon number decays with rate $A$ towards its stationary value of
$\nu$.
\end{list}
Thus the meaning of $A$ and $\nu$ is quite clear.

The relation between the temperature of the resonator and the number of
thermal photons deserves a little attention. With the numbers of
Eqs.~(\ref{freqnumb}) and (\ref{enernumb}) we find the entries in this
table:
\begin{center}
\begin{tabular}{c|c|l}
$T$&$\nu$&remarks\\
\hline
1.5\,K&1.96&early experiments\\
0.5\,K&0.15&typical values\\
80\,mK&$2.5\times10^{-6}$&achieved
\end{tabular}
\end{center}
Please observe that in the microwave domain thermal photons are usually
present in considerable amounts. This situation is quite different from that
of the optical domain in which thermal photons are irrelevant.

Another remark concerns the connection between the decay constant $A$ and the
quality factor $Q$ of the resonator. Per definition the energy stored
decreases to $1/e$-th of its initial value while the field undergoes $Q$
cycles,
\begin{equation}
e^{-At}=e^{-1}\quad \mbox{for}\quad \omega t=2\pi Q\;:\ %
A=\left.\frac{\omega}{2\pi}\right/Q\,.
\end{equation}
In the Garching micromaser experiments one has
$\omega\cong2\pi\times21.5\,\mbox{GHz}$, and quality factors of
$Q\cong10^9\cdots10^{10}$ have been achieved, so that the photon life time
$1/A$ can be as large as a few hundred microseconds.

\SEC{One-atom maser: Steady state. General matters}

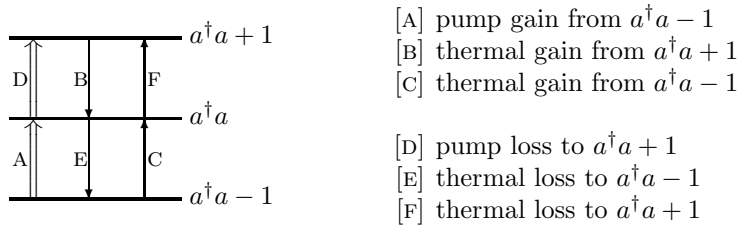
\begin{figure}[tb]
\begin{picture}(390,115)(-30,10)
\put(25,50){$\Bigg\Uparrow$}
\put(25,80.5){$\Bigg\Uparrow$}
\put(71,37.5){\vector(0,1){31}}
\put(71,68){\vector(0,1){31}}
\put(50,68){\vector(0,-1){31}}
\put(50,98.5){\vector(0,-1){31}}
\thicklines
\put(20,37.5){\line(1,0){65}}
\put(20,68){\line(1,0){65}}
\put(20,98.5){\line(1,0){65}}
\thinlines
\put(21,80.5){\footnotesize\sc d}
\put(44,80.5){\footnotesize\sc b}
\put(72,80.5){\footnotesize\sc f}
\put(21,50){\footnotesize\sc a}
\put(44,50){\footnotesize\sc e}
\put(72,50){\footnotesize\sc c}
\put(88,35.5){\footnotesize$a^{\dag}a-1$}
\put(88,66){\footnotesize$a^{\dag}a$}
\put(88,96.5){\footnotesize$a^{\dag}a+1$}
\put(160,66){\parbox[c]{160pt}{\footnotesize\tabcolsep=2pt
\begin{tabular}{rl}
[{\sc a}]&pump gain from $a^{\dag}a-1$\\{}
[{\sc b}]&thermal gain from $a^{\dag}a+1$\\{}
[{\sc c}]&thermal gain from $a^{\dag}a-1$\\ & \\{}
[{\sc d}]&pump loss to $a^{\dag}a+1$\\{}
[{\sc e}]&thermal loss to $a^{\dag}a-1$\\{}
[{\sc f}]&thermal loss to $a^{\dag}a+1$
\end{tabular}
}}
\end{picture}
\caption[Gain and loss]{\label{gainloss}\small\baselineskip=12pt
Probability fluxes representing the gain and loss terms in the OAM master
equation for a diagonal photon state. The labeling agrees with the one in
Eq.~(\ref{diagME}). \hrulefill}
\end{figure}

The OAM master equation (\ref{OAMmaster}) with (\ref{MEfree}),
(\ref{MEgain}), and (\ref{MEloss}) does not contain any terms that are
sensitive to the phase of the photon field.\footnote{In more technical terms
this statement says that (\ref{OAMmaster}) is invariant under the unitary
transformation $\rho\to U^{\dag}\rho U$ with $U=\exp(i\chi\N)$ where $\chi$
is an arbitrary (constant) phase angle.} Another way of putting this states
that the diagonals of the number-state matrix of $\rho$ are dynamically
uncoupled. Therefore, a state $\rho$ which at one time is diagonal in this
sense, stays diagonal for all times. A diagonal state is represented by a
state operator $\rho$ that depends solely on the product $\N$, the photon
number operator, but not on the ladder operators $a$ and $a^{\dag}$
individually: $\rho=\rho(\N)$. The numbers $\rho(n)=\bra{n}\rho(\N)\ket{n}$
are the probabilities to find $n$ photons in the resonator.

For such a diagonal state, the master equation (\ref{OAMmaster}) is
particularly transparent. It constitutes a third central result of micromaser
theory:
\begin{equation}\label{diagME}
\fbox{\parbox{280pt}{
\begin{center}
\hspace*{10pt}Master equation for diagonal states\hspace*{10pt}\\
of the one-atom maser:
\begin{displaymath}
\renewcommand{\arraystretch}{1.4}\arraycolsep=0pt
\begin{array}{rrcl}
\mbox{\Large$\partt$}\rho_t(\N)=&
r\sin^2\left[\varphi(\N)^{1/2}\right]\rho_t(\N-1)&\quad&[\mbox{\sc a}]\\
&+A(\nu+1)(\N+1)\rho_t(\N+1)&&[\mbox{\sc b}]\\
&+A\nu\N\rho_t(\N-1)&&[\mbox{\sc c}]\rule[-15pt]{0pt}{5pt}\\ 
&-r\sin^2\left[\varphi(\N+1)^{1/2}\right]\rho_t(\N)&&[\mbox{\sc d}]\\
&-A(\nu+1)\N\rho_t(\N)&&[\mbox{\sc e}]\\
&-A\nu(\N+1)\rho_t(\N)&&[\mbox{\sc f}]
\end{array}
\end{displaymath}
\end{center}
}}
\end{equation}
The six terms on the right-hand side possess obvious meanings. These are
stated in the probability flux diagram of Fig.~\ref{gainloss}.

In the steady state $\rho^{\stext{(SS)}}$ of (\ref{diagME}) --- which is, of
course, also the steady state of (\ref{OAMmaster}) --- the gains and losses
of $\rho(\N)$ have to add up to zero. A rearrangement of the right-hand-side
terms in (\ref{diagME}),
\begin{nareqns}{rll}\label{3terms}
r\sin^2\left[\varphi(\N)^{1/2}\right]\rho^{\stext{(SS)}}(\N-1)
&\qquad&[\mbox{\sc a}]\\
+A\nu\N\rho^{\stext{(SS)}}(\N-1)&&[\mbox{\sc c}]\\ 
-A(\nu+1)\N\rho^{\stext{(SS)}}(\N)&&[\mbox{\sc e}]\rule[-15pt]{0pt}{5pt}\\
=r\sin^2\left[\varphi(\N+1)^{1/2}\right]\rho^{\stext{(SS)}}(\N)
&&[\mbox{\sc d}]\\
A\nu(\N+1)\rho^{\stext{(SS)}}(\N)&&[\mbox{\sc f}]\\
-A(\nu+1)(\N+1)\rho^{\stext{(SS)}}(\N+1)&\,,&[\mbox{\sc b}]
\end{nareqns}%
combines them into two groups of three. Now, observe that the replacement
$\N\to\N+1$ turns the [{\sc a,c,e}] group into the [{\sc d,f,b}] group. By
induction we therefore conclude that the value of the [{\sc a,c,e}] group
is the same for each value of $\N$. But for $\N=0$ it is zero. So we arrive
at the two-term recurrence relation
\begin{equation}\label{2terms}
A(\nu+1)\N\rho^{\stext{(SS)}}(\N)=
\left(r\sin^2[\varphi(\N)^{1/2}]
+A\nu\N\right)\rho^{\stext{(SS)}}(\N-1)\,,
\end{equation}
which is equivalent to the three-term relation (\ref{3terms}) but
considerably simpler. 

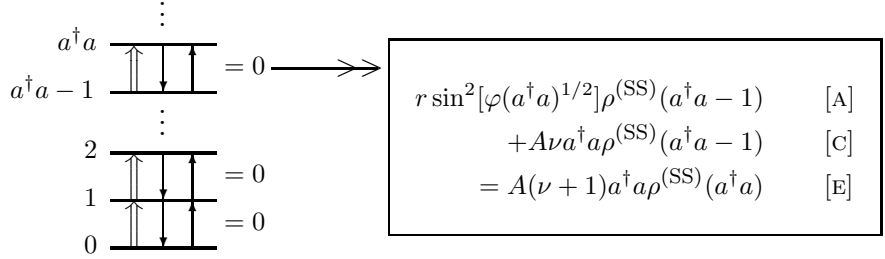
\begin{figure}[tb]
\begin{picture}(390,130)(25,0)
\put(156,84){\line(1,0){40}}    
\put(180,81.1){$>>$}
\put(200,24){\fbox{\parbox[b]{180pt}{\footnotesize
\begin{eqnarray*}
r\sin^2[\varphi(\N)^{1/2}]\rho^{\mbox{\scriptsize(SS)}}(a^{\dag}a-1)
&\;& [\mbox{\sc a}]\\
+A\nu a^{\dag}a\rho^{\mbox{\scriptsize(SS)}}(a^{\dag}a-1)&& [\mbox{\sc c}]\\
=A(\nu+1)a^{\dag}a\rho^{\mbox{\scriptsize(SS)}}(a^{\dag}a)&& [\mbox{\sc e}]
\end{eqnarray*}
}}}
\put(138,23){\footnotesize $=0$}
\put(138,41){\footnotesize $=0$}
\put(138,82){\footnotesize $=0$}
\put(85,14){\footnotesize 0}
\put(85,32){\footnotesize 1}
\put(85,50){\footnotesize 2}
\put(57,73){\footnotesize $a^{\dag}a-1$}
\put(75,91){\footnotesize $a^{\dag}a$}
\put(113,100){$\vdots$}
\put(113,59){$\vdots$}
\put(100,22){$\Big\Uparrow$}
\put(115,34){\vector(0,-1){18}}
\put(126,16){\vector(0,1){18}}
\put(100,40){$\Big\Uparrow$}
\put(115,52){\vector(0,-1){18}}
\put(126,34){\vector(0,1){18}}
\put(100,81){$\Big\Uparrow$}
\put(115,93){\vector(0,-1){18}}
\put(126,75){\vector(0,1){18}}
\thicklines
\put(95,16){\line(1,0){40}}
\put(95,34){\line(1,0){40}}
\put(95,52){\line(1,0){40}}
\put(95,75){\line(1,0){40}}
\put(95,93){\line(1,0){40}}
\end{picture}
\caption[Detailed balance]{\label{detbal}\small\baselineskip=12pt
In steady state the probability fluxes obey a detailed balance between
neighboring photon numbers. The terms are labeled the same way as in
Eq.~(\ref{diagME}) and in Fig.~\ref{gainloss}. \hrulefill}
\end{figure}

The transition from (\ref{3terms}) to (\ref{2terms}) recognizes the detailed
balance in steady state, illustrated in Fig.~\ref{detbal}.
It comes about because the photon number ladder ends at the $\N=0$ rung.

The recurrence relation (\ref{2terms}) can be solved immediately, and we
obtain a fourth central result of micromaser theory:
\begin{equation}\label{OAMss}
\fbox{\parbox{310pt}{
\begin{center}
\parbox{220pt}{The steady state of the one-atom maser,
pumped by resonant $\UP$ atoms, is given by}
\begin{displaymath}
\rho^{\stext{(SS)}}(\N)=\rho^{\stext{(SS)}}(0)\prod_{n=1}^{\mbox{$\N$}}%
\left[\frac{\nu}{\nu+1}+\frac{r/A}{\nu+1}\frac{\sin^2(\varphi\sqrt{n})}{n}%
\right]\,,
\end{displaymath}
\parbox{220pt}{where the value of $\rho^{\stext{(SS)}}(0)$ is
determined\break by the normalization of $\rho^{\stext{(SS)}}$ to unit
trace.}
\end{center}
}}
\end{equation}
We note that this steady state is determined uniquely by three numbers: the
thermal photon number $\nu$, the accumulated Rabi angle $\varphi$, and the
ratio $r/A$ of the pump and the decay rates. This ratio tells us how many
pump atoms traverse the resonator (on average) during one photon life time.
Therefore, we call $r/A$ the effective pump rate.\footnote{In the micromaser
literature the symbol $N_{\stext{ex}}$ is frequently used as an abbreviation
for $r/A$ where the subscript `ex' is a reminder of `excited.' We do not
adopt this custom.}

A simple, but necessary, check of consistency is provided by the
consideration of the $r=0$ case, the case of an unpumped resonator. The
steady state of (\ref{OAMss}) is then the thermal state $\rho_{\stext{th}}$
of (\ref{thermal}), as it must be.

\SEC{One-atom maser: Steady state. Examples}

\savebox{\texta}{\footnotesize $\langle\N\rangle$}
\savebox{\textb}{\footnotesize Fano factor}
\begin{figure}[p]
\begin{picture}(390,470)(30,0)
\put(100,15){\epsfysize=220pt\epsffile[85 45 547 411]{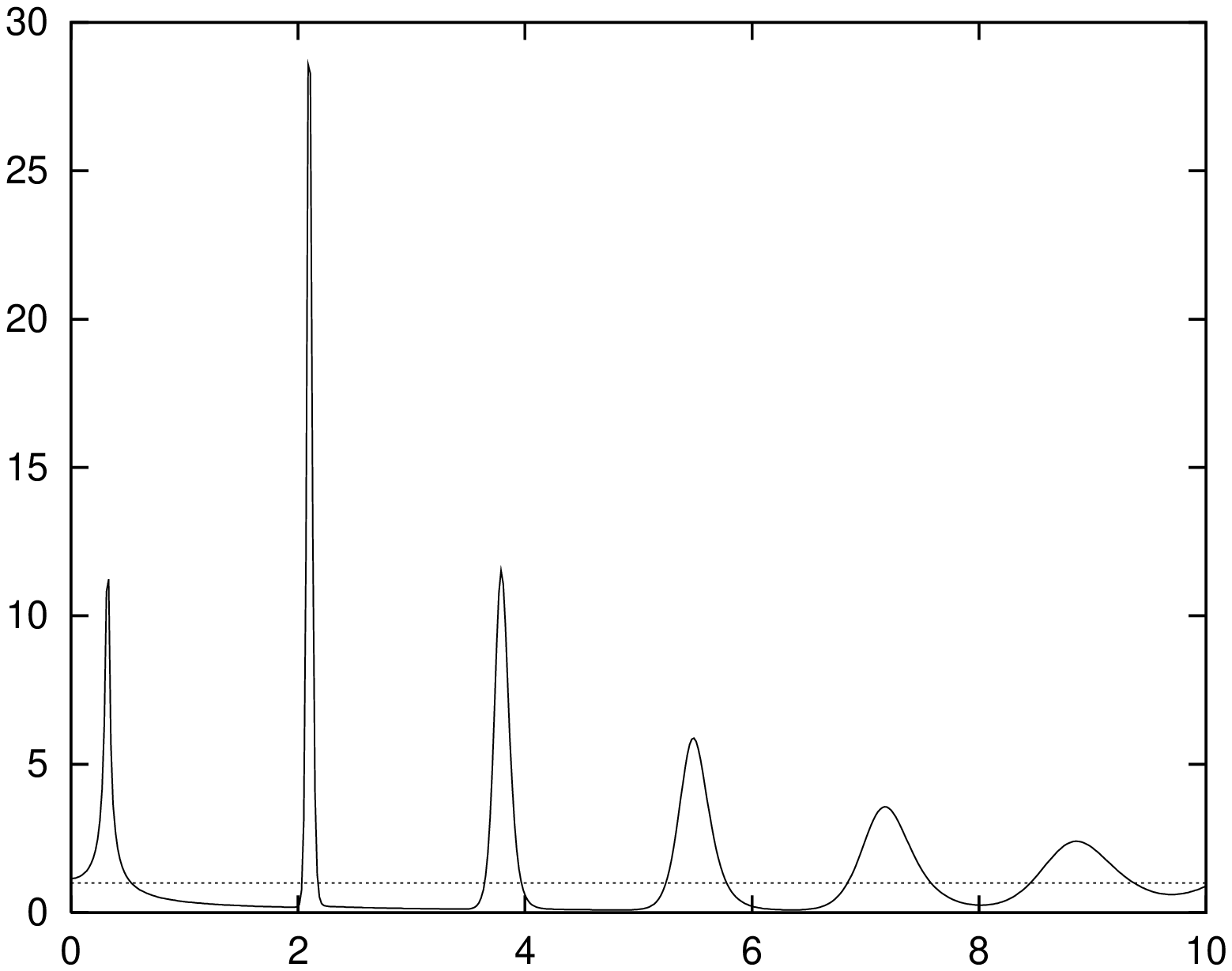}}
\put(100,250){\epsfysize=220pt\epsffile[85 45 547 411]{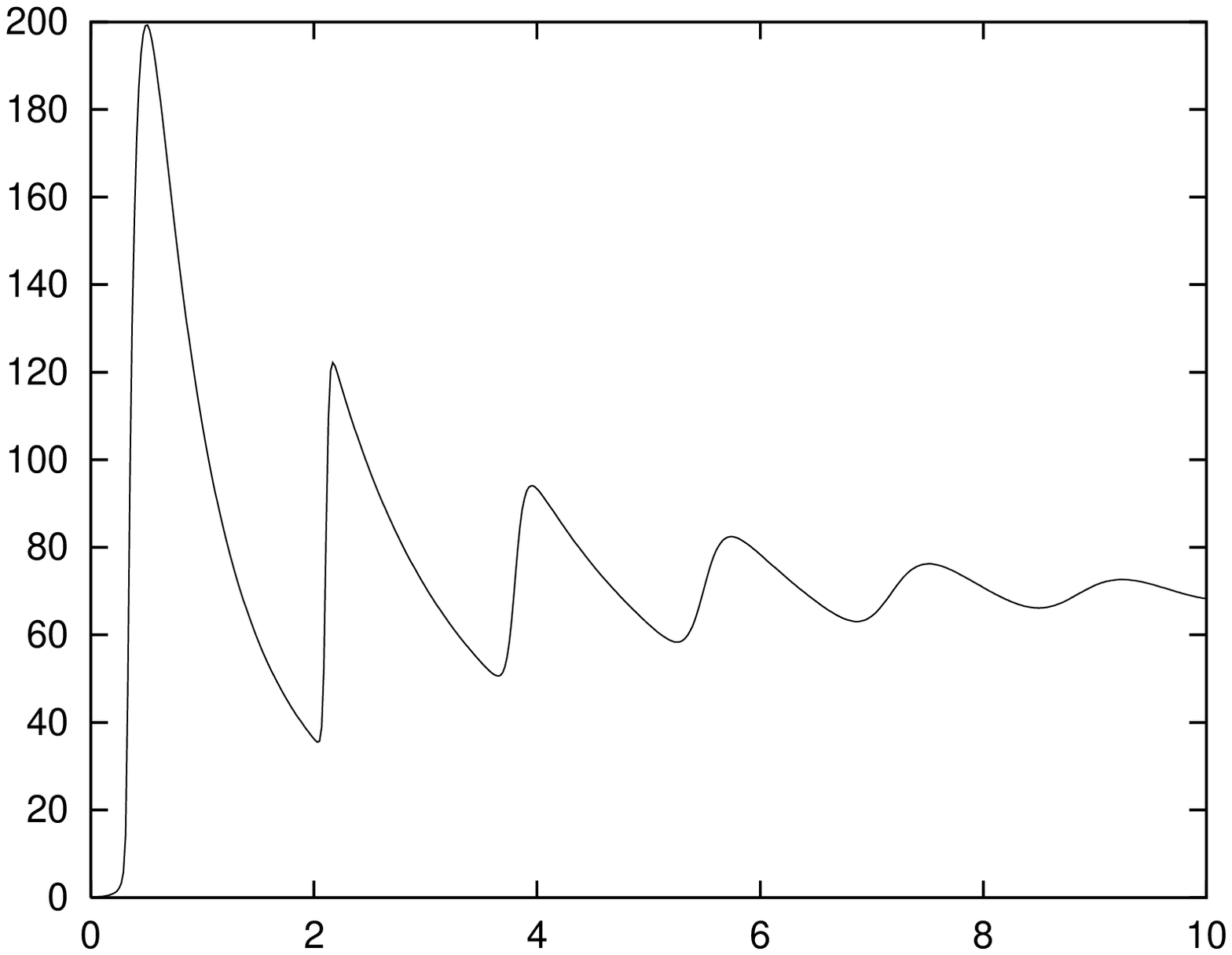}}
\put(234,10){\footnotesize $\theta/\pi$}
\put(234,245){\footnotesize $\theta/\pi$}
\put(80,350){\rotl{\texta}}
\put(85,105){\rotl{\textb}}
\end{picture}
\caption[Steady state for $r/A=200,\nu=0.15$]{\label{OAM-ss-200-0.15}
Mean photon number and Fano factor in the OAM steady state for an effective
rate of $r/A=200$ and a thermal photon number of $\nu=0.15$, as a function
of the pump parameter $\theta$ in the range from $\theta=0$ to
$\theta=10\pi$.}
\end{figure}
\savebox{\textb}{\small Fano factor}
\begin{figure}[t]
\begin{picture}(390,210)(70,0)
\put(94,15){\epsfxsize=360pt\epsffile[85 45 547 411]{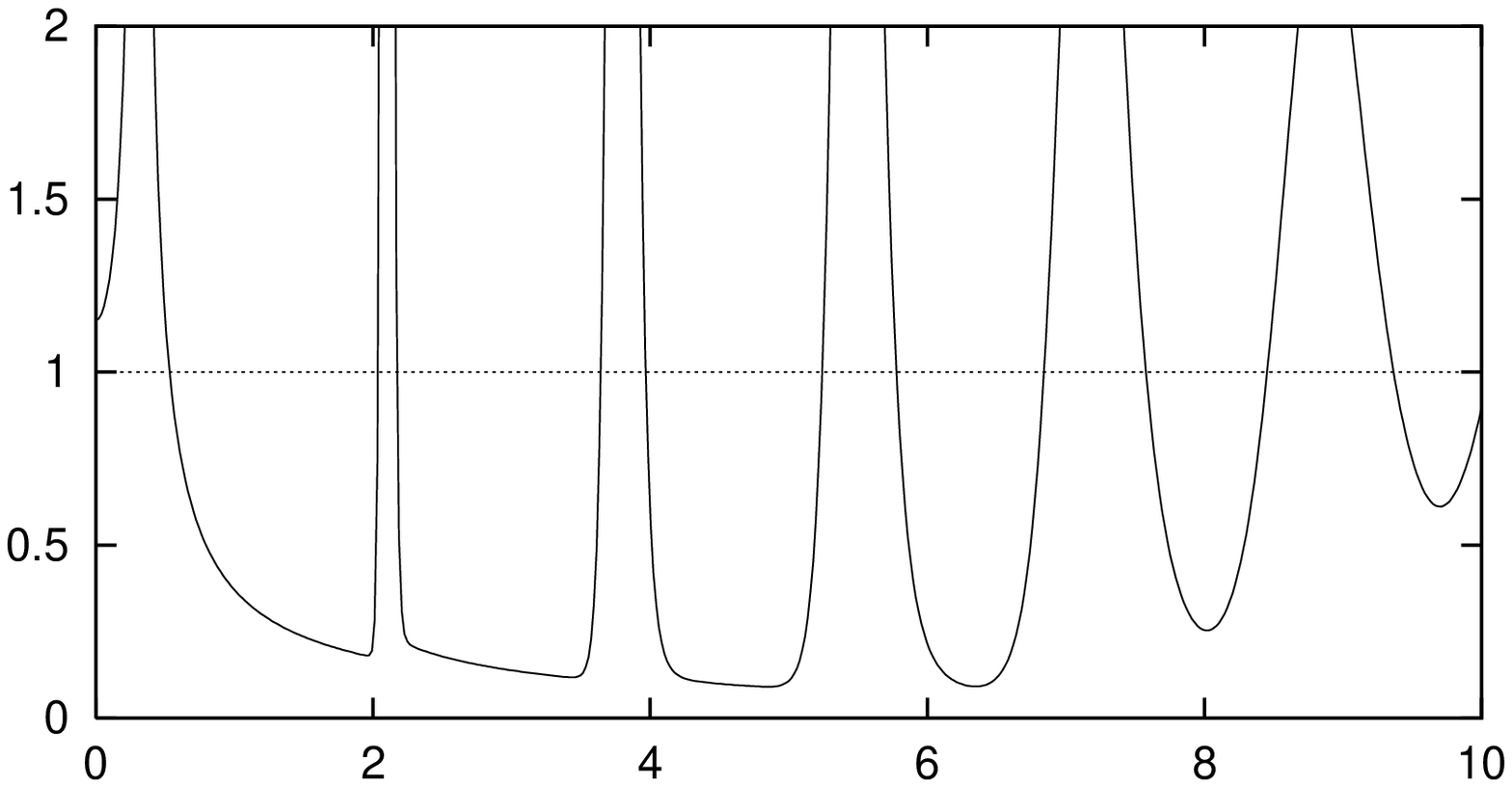}}
\put(280,10){\small $\theta/\pi$}
\put(85,93){\rotl{\textb}}
\end{picture}
\caption[Steady state for $r/A=200,\nu=0.15$]{\label{OAM-ss-200-0.15-a}
Fano factor in the OAM steady state for an effective rate of $r/A=200$ and
a thermal photon number of $\nu=0.15$, as a function of the pump parameter
in the range $0\leq\theta/\pi\leq10$. \hrulefill}
\end{figure}

For fixed values of the effective pump rate $r/A$ and the thermal photon
number $\nu$, the steady state of the OAM is traditionally studied as a
function of the so-called `pump parameter' $\theta\equiv\varphi\sqrt{r/A}$.
Why this scaling of the Rabi angle appears convenient will become plausible
later.

For a certain OAM steady state $\rho$ --- for simplicity, we drop the
superscript of $\rho^{\stext{(SS)}}$ until we start considering the time
dependence of $\rho$ again --- the fundamental statistical quantities of
interest are mean the photon number $\langle\N\rangle=\mbox{tr}\{\N\rho\}$
and its normalized variance,
\begin{equation}\label{Fano}
\frac{\langle(\N)^2\rangle-\langle\N\rangle^2}{\langle\N\rangle}.
\end{equation}
This quantity (or its square root) is known under various names in the
lit\-era\-ture,\footnote{These names derive from a 1947 paper by
Fano~\cite{Fano47} and a 1979 paper by Mandel~\cite{Man79}, I~believe.} such
as `Fano factor,' or `Mandel's Q parameter,' or `Fano-Mandel measure.' We
prefer the first one in this list. The normalization in (\ref{Fano}) is to
the value of a Poissonian distribution (as in a coherent state of the
harmonic oscillator or a classical state of the radiation field),
\begin{equation}\label{poisson}
\rho(n)_{\Big\vert_{\stext{Poisson}}}
=\frac{1}{n!}\langle\N\rangle^n\exp(-\langle\N\rangle)\,,
\end{equation}
for which $\langle(\N)^2\rangle-\langle\N\rangle^2=\langle\N\rangle$. With
this normalization comes the jargon of calling a state $\rho$
`superpoissonian' if its Fano factor exceeds unity, and `subpoissonian' if it
is less than unity. An example for a superpoissonian state is the thermal
state of Eq.~(\ref{thermal}); its Fano factor equals $\nu+1$.

In Fig.~\ref{OAM-ss-200-0.15} the dependence of the photon number and the
Fano factor on the pump parameter $\theta$ is shown for an effective pump
rate of $r/A=200$ and $\nu=0.15$ thermal photons. 
As $\theta$ increases from $\theta=0$, we see an initial rapid growth of
both the photon number and the Fano factor. Both numbers reach a maximum and
then decrease toward a minimum near $\theta=2\pi$. This pattern repeats
itself roughly with a $\theta$ period of $2\pi$. 

The maximal value of $\langle\N\rangle$, reached near $\theta=\pi/2$, is very
close to 200, the value of the effective pump rate. Thus, for this pump
parameter the interaction time is just right to ensure that almost all atoms
emit a photon into the resonator. At the minimum near $\theta=2\pi$, by
contrast, a large fraction of the atoms do not emit. The interaction time is
longer, so that the atoms have a good change to undergo a complete Rabi cycle
$\ket{\UP}\to\ket{\DN}\to\ket{\UP}$ and leave the resonator in the excited
state.

The bottom part of the plot of the Fano factor in Fig.~\ref{OAM-ss-200-0.15}
is enlarged in Fig.~\ref{OAM-ss-200-0.15-a}. We observe that there are large
$\theta$ intervals in which the Fano factor is less than one, thus indicating
a subpoissonian state of the photon field. It is clear that such states can
be realized experimentally, since a very precise control of the parameters
$r/A$, $\nu$, and $\theta$ is not necessary to stay inside the appropriate
parameter range.

\savebox{\texta}{\footnotesize $\rho(n)$}
\savebox{\textb}{\footnotesize $\rho(n)$}
\savebox{\textc}{\footnotesize $\rho(n)$}
\savebox{\textd}{\footnotesize $\rho(n)$}
\begin{figure}[tbp]
\begin{picture}(390,305)(0,0)
\put(20,170){\epsfxsize=170pt\epsffile[85 45 547 411]{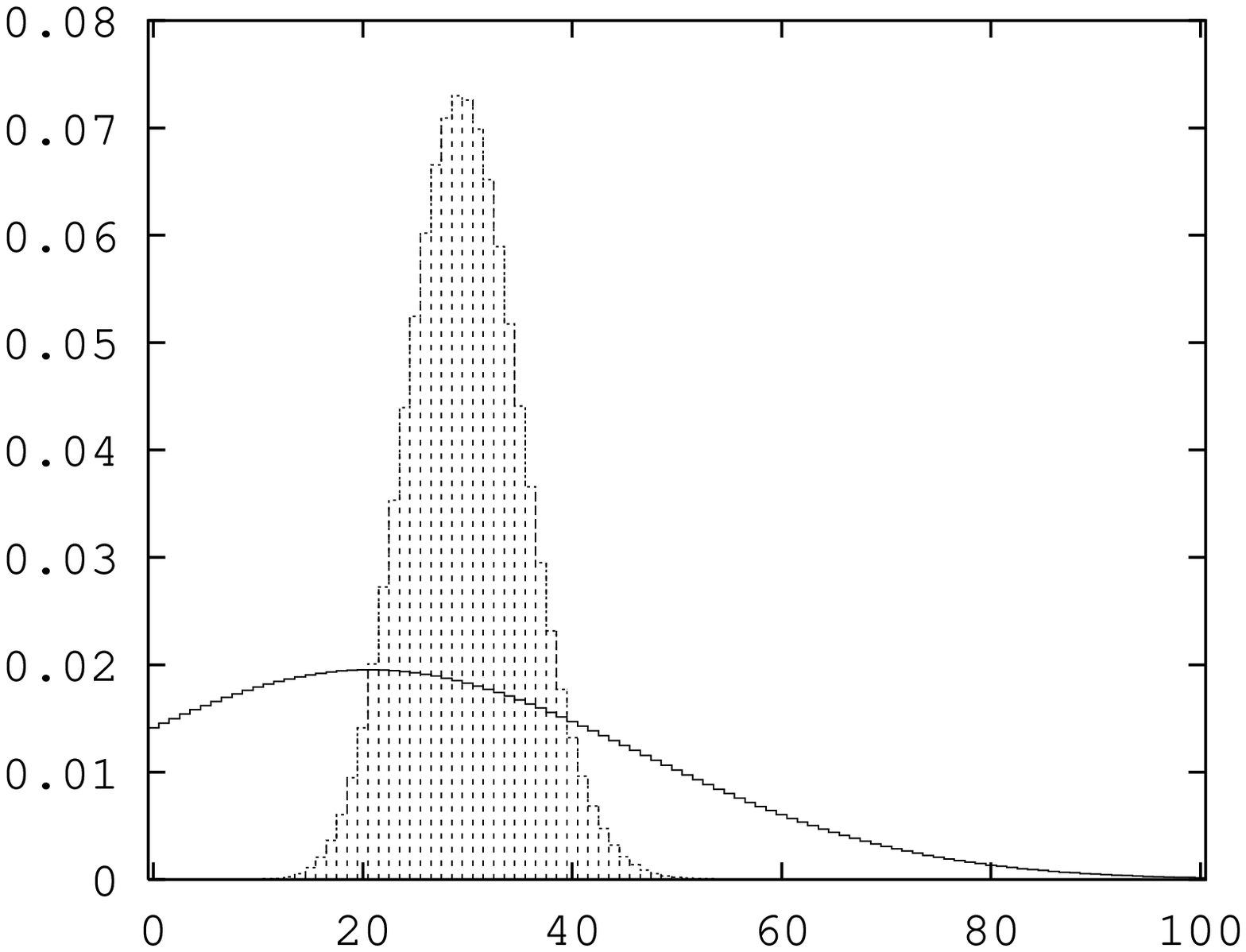}}
\put(220,170){\epsfxsize=170pt\epsffile[85 45 547 411]{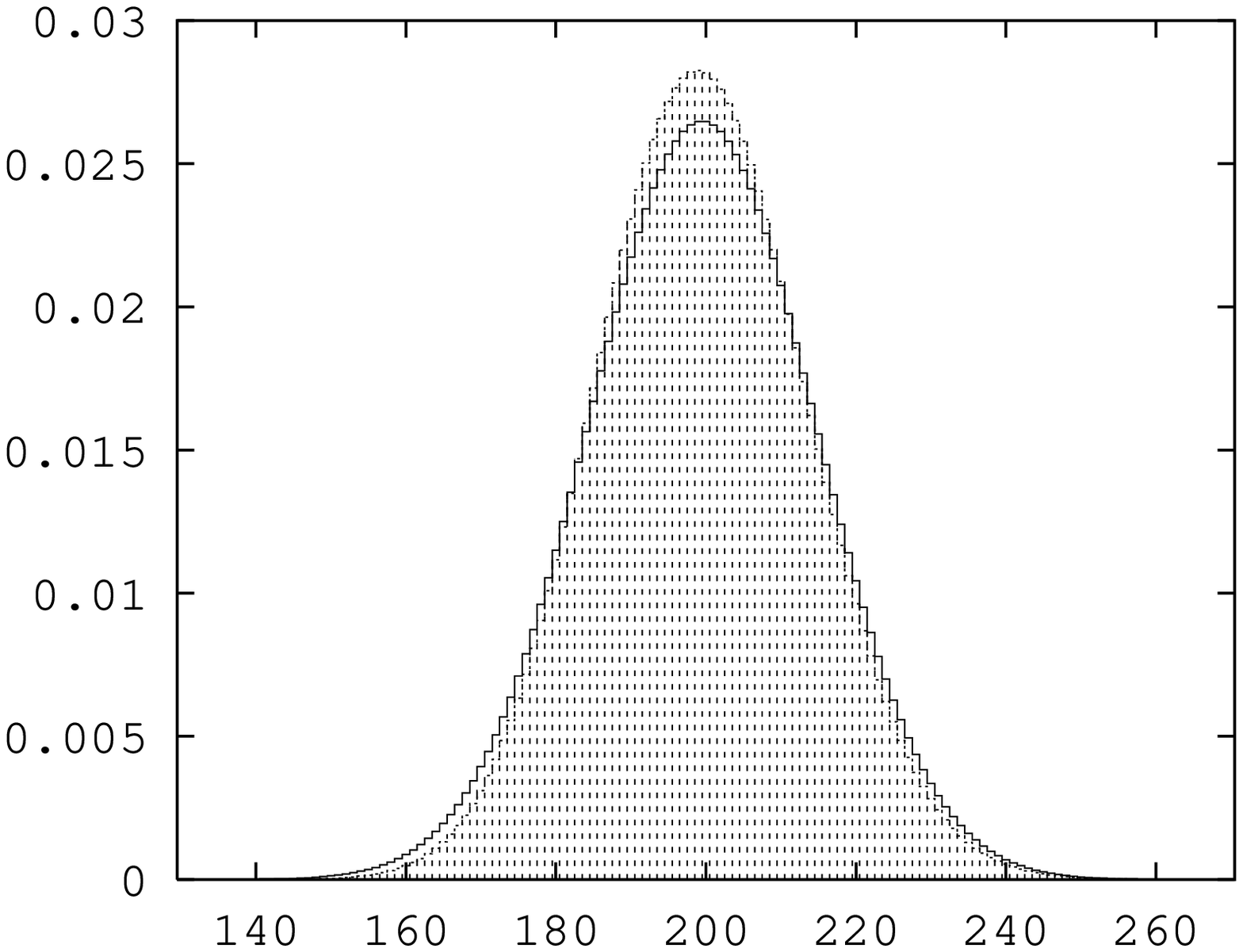}}
\put(20,15){\epsfxsize=170pt\epsffile[85 45 547 411]{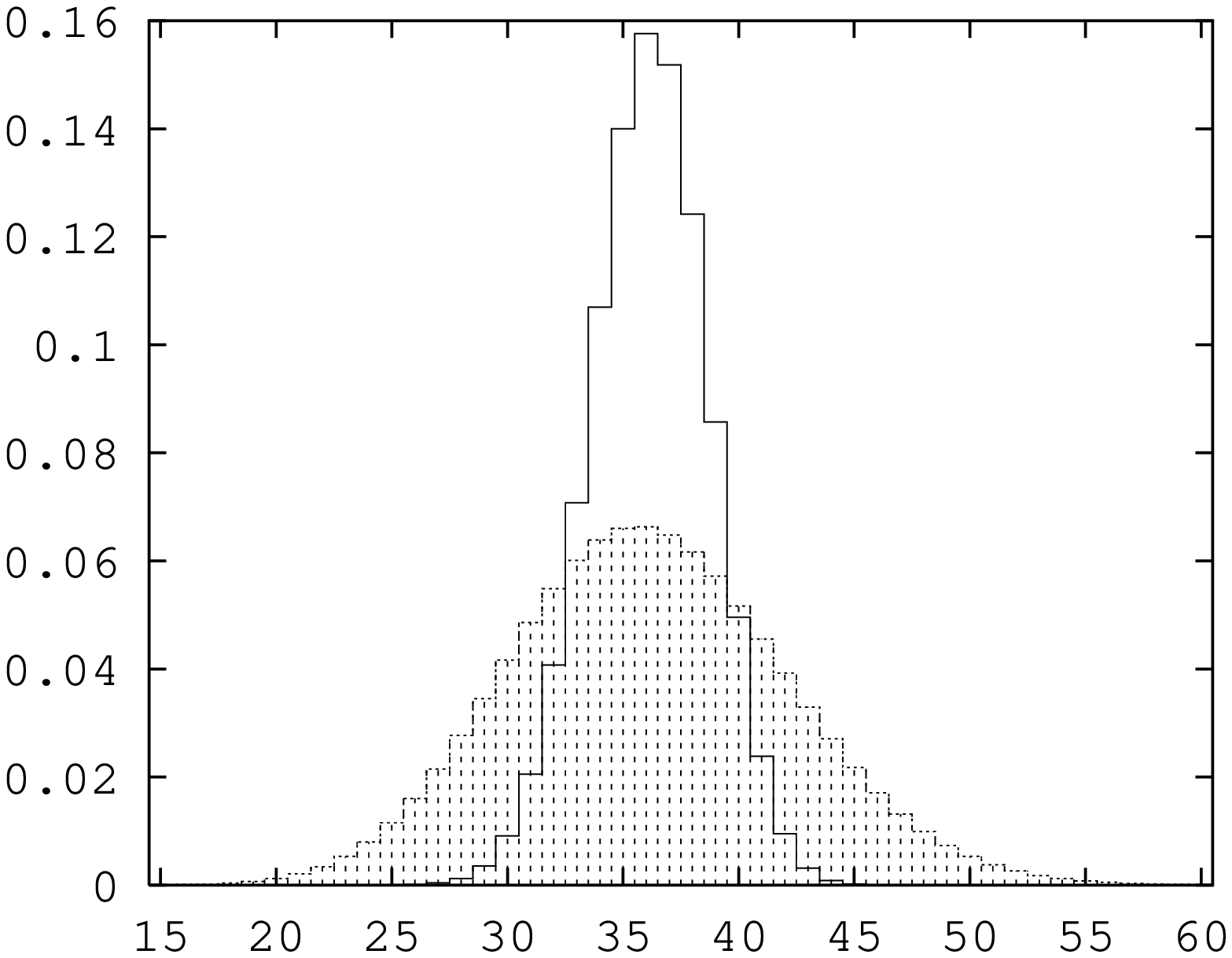}}
\put(220,15){\epsfxsize=170pt\epsffile[85 45 547 411]{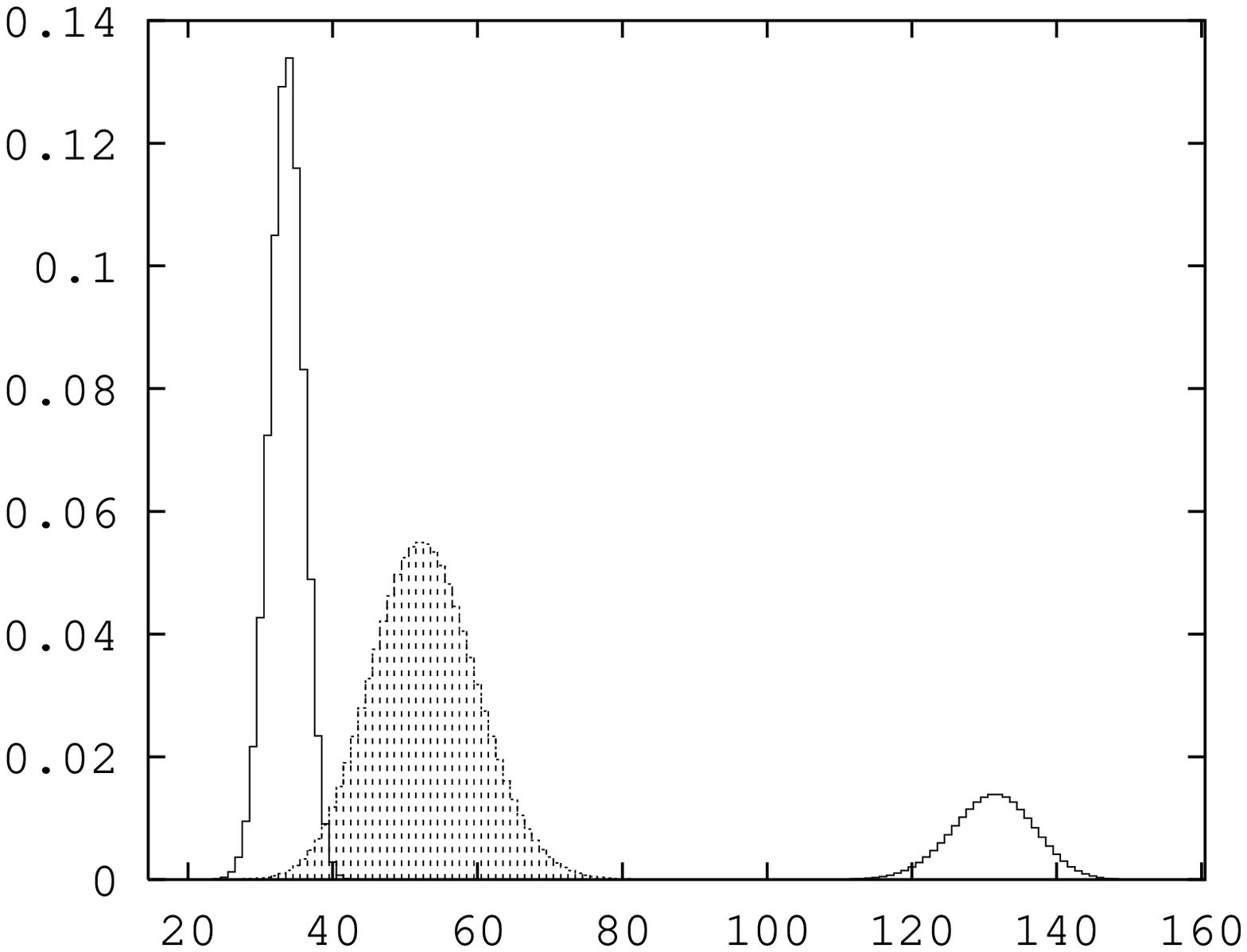}}
\put(5,235){\rotl{\texta}}
\put(202,235){\rotl{\textb}}
\put(5,80){\rotl{\textc}}
\put(205,80){\rotl{\textd}}
\put(108,5){\footnotesize$n$}
\put(306,5){\footnotesize$n$}
\put(108,160){\footnotesize$n$}
\put(306,160){\footnotesize$n$}
\put(160,280){\footnotesize (a)}
\put(360,280){\footnotesize (b)}
\put(160,125){\footnotesize (c)}
\put(360,125){\footnotesize (d)}
\end{picture}
\caption[Histograms]{\label{histos}
Histograms for the photon number distribution in the OAM steady state for an
effective rate of $r/A=200$ and a thermal photon number of $\nu=0.15$. The
plots are for (a)~$\theta=0.32442\pi$ (i.\,e., $\varphi=4.129^{\circ}$),
(b)~$\theta=\pi/2$ (i.\,e., $\varphi=6.364^{\circ}$), (c)~$\theta=2\pi$
(i.\,e., $\varphi=25.46^{\circ}$), and (d)~$\theta=2.09\pi$ (i.\,e.,
$\varphi=26.60^{\circ}$). The solid line shows the actual distribution. The
broken line with the shaded area underneath shows the corresponding Poisson
distribution with the same mean photon number. The respective mean photon
numbers $\langle\N\rangle$ are (a)~30.14, (b)~199.44, (c)~36.25, and
(d)~52.72. The Fano factors equal (a)~12.56, (b)~1.153, (c)~0.214, and
(d)~28.68.}
\end{figure}

A look at the histograms in Fig.~\ref{histos} might be helpful in
understanding why the Fano factor covers such a large range of values.
For reference each histogram is compared with a corresponding Poissonian one
with the same mean number of photons. Whereas we have an almost poissonian
distribution for $\theta=\pi/2$ (at the maximum of the $\langle\N\rangle$),
the distribution for $\theta=0.32442\pi$ is much broader than the Poissonian
one, and that for $\theta=2\pi$ is much narrower. In these cases it is easily
understood that the respective Fano factors are close to, much larger than,
or much smaller than unity. A different situation is observed for
$\theta=2.09$ where, according to Fig.~\ref{OAM-ss-200-0.15}, the Fano factor
is particularly large. In Fig.~\ref{histos} we recognize the reason for that:
there are two narrow peaks --- each subpoissonianly narrow --- separated by
many photon numbers $n$. As a consequence, the Fano factor, which is a global
property of the distribution, is enormous.

\savebox{\texta}{\small $\rho(n)$}
\begin{figure}[tb]
\begin{picture}(390,215)(70,0)
\put(100,15){\epsfxsize=360pt\epsffile[85 45 540 303]{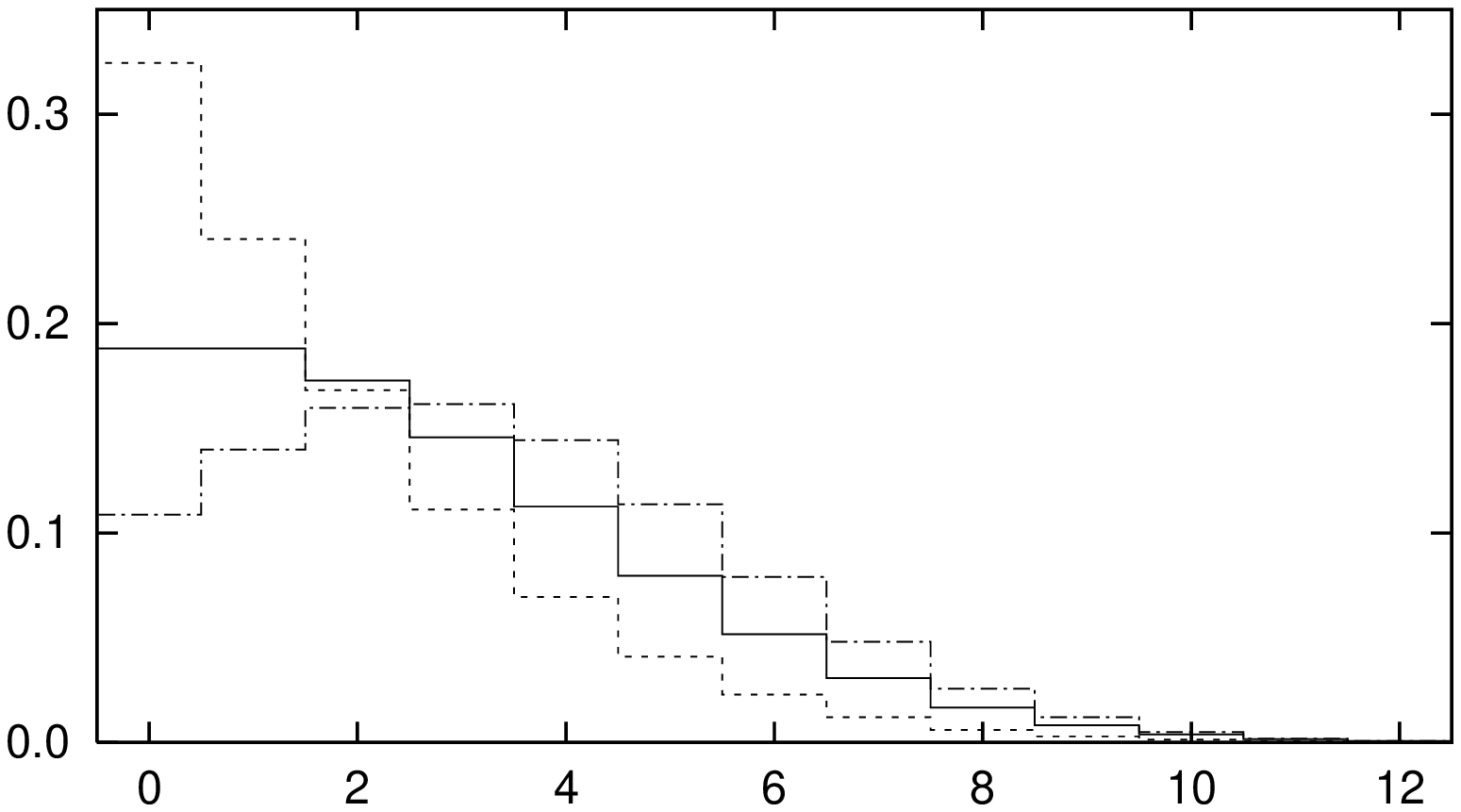}}
\put(260,10){\small $n$}
\put(80,110){\rotl{\texta}}
\end{picture}
\caption[Maser threshold]{\label{OAMthresh}
Histograms for the photon number distribution in the OAM steady state for
an effective pump rate of $r/A=4$ and $\nu=0.1$ thermal photons. The plot
shows the histograms for three different Rabi angle: $\varphi=30^{\circ}$
(line $\frac{\hspace*{2em}}{\hspace*{2em}}$, at threshold),
$\varphi=25^{\circ}$ (line $---$, below threshold), and $\varphi=35^{\circ}$
(line $-\cdot-\cdot-$, above threshold). \hrulefill}
\end{figure}

The notion of the `maser threshold' is also most easily understood in the
context of such histograms.
The thermal state (\ref{thermal}) is surely below threshold by any
definition. Likewise the possonian state (\ref{poisson}), which is so
typical for the output of a laser, is certainly above threshold.
A striking difference between the two is that the maximum of the poissonian
state is at a large photon number $n$, whereas in the thermal state the
largest probability is the one for zero photons. This invites the following
criterion based on the relative size of $\rho(0)$, the probability for no
photons, and $\rho(1)$, that for one photon:
\begin{eqns}{rll}
\rho(0)>\rho(1)&:\quad&\mbox{below threshold,}\\
\rho(0)=\rho(1)&:\quad&\mbox{at threshold,}\\
\rho(0)<\rho(1)&:\quad&\mbox{above threshold.}
\end{eqns}%
In view of the recurrence relation (\ref{2terms}), the explicit OAM version
reads
\begin{equation}\label{bel-at-abo}
\frac{r}{A}\sin^2(\varphi)\left\{
\begin{array}{c}
< \\ = \\ >
\end{array}
\right\} 1\,:\quad\left\{
\begin{array}{c}
\mbox{below} \\ \mbox{at} \\ \mbox{above}
\end{array}
\right\}\ \mbox{threshold.}
\end{equation}
In particular, if the effective rate $r/A$ is large, the threshold value of
$\varphi$ is so small that $\sin^2(\varphi)$ can be replaced by $\varphi^2$,
and then we have simply $\theta=1$ at the maser threshold. This is one reason
why the pump parameter $\theta$ is sometimes preferred over the unscaled Rabi
angle $\varphi$. These matters are illustrated in Fig.~\ref{OAMthresh} where
the value $r/A=4$ is chosen for the effective rate, so that
$\varphi=30^{\circ}$ holds at the threshold. We note that the thermal photon
number $\nu$ plays no role in (\ref{bel-at-abo}).

\begin{figure}[tb]
\begin{picture}(390,140)(85,0)
\put(100,15){\epsfxsize=360pt\epsffile[99 50 540 205]{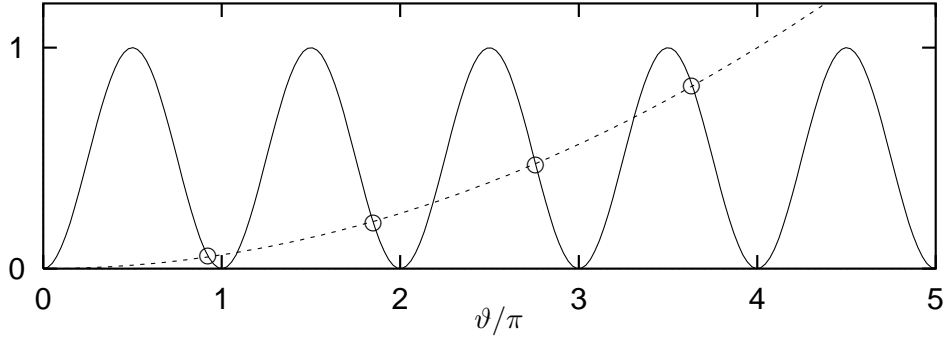}}
\put(278,10){\small $\vartheta/\pi$}
\put(177,37){\circle{6}}
\put(239.5,49.5){\circle{6}}
\put(301,71.5){\circle{6}}
\put(360,101.5){\circle{6}}
\end{picture}
\caption[Maxima and minima]{\label{maximin}
The two functions of (\ref{regions}) are plotted for 
$0\leq \vartheta/\pi\leq5$. The pump parameter $\theta$ has the value of
$4\pi$. The intersections marked with a little circle $\circ$ determine the
maxima in the photon number distribution, the other intersections determine
the minima. \hrulefill}
\end{figure}
\savebox{\texta}{\small $\rho(n)$}
\begin{figure}[tb]
\begin{picture}(390,215)(-8,0)
\put(20,15){\epsfxsize=360pt\epsffile[85 48 553 408]{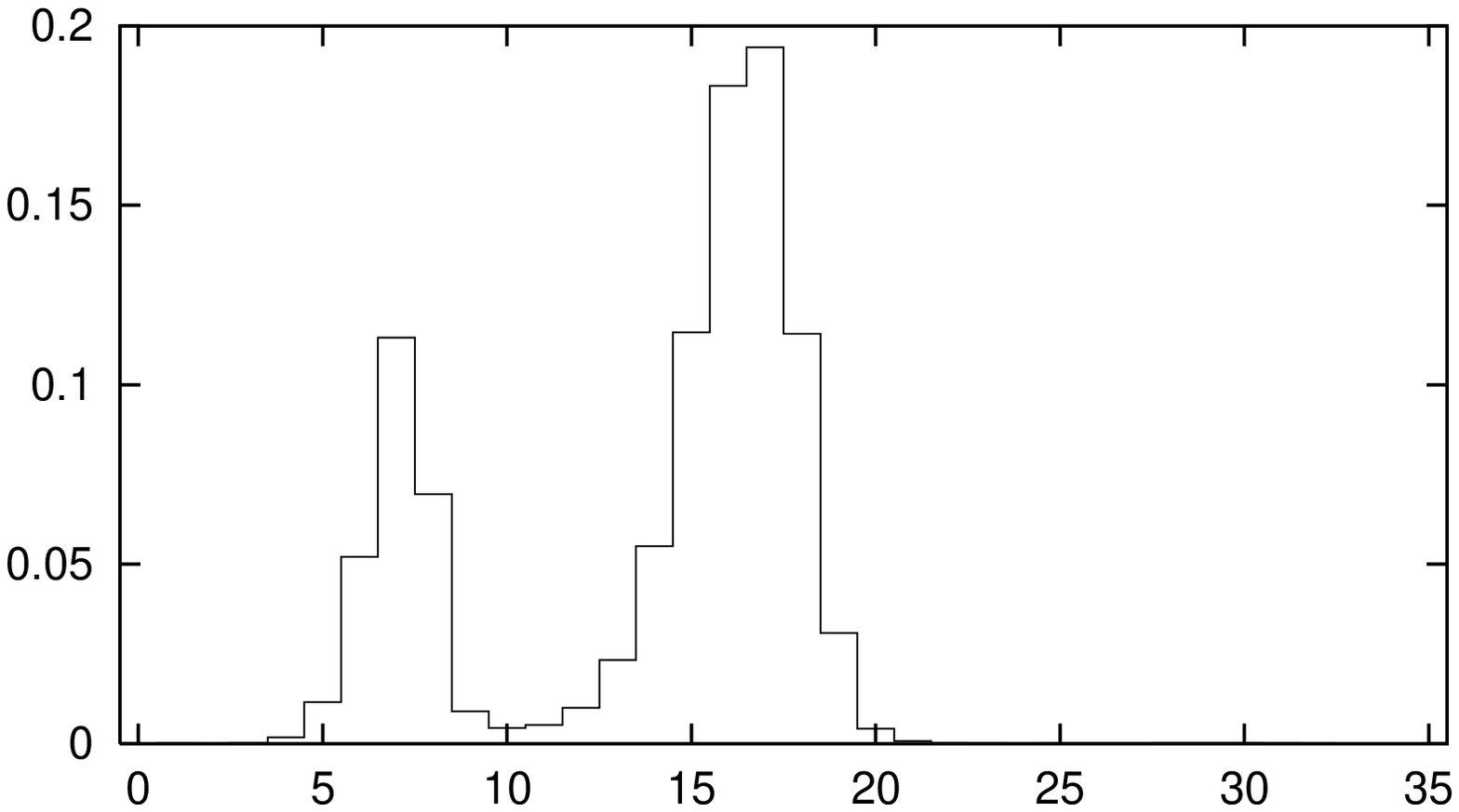}}
\put(20,15){\epsfxsize=360pt\epsffile[85 48 553 408]{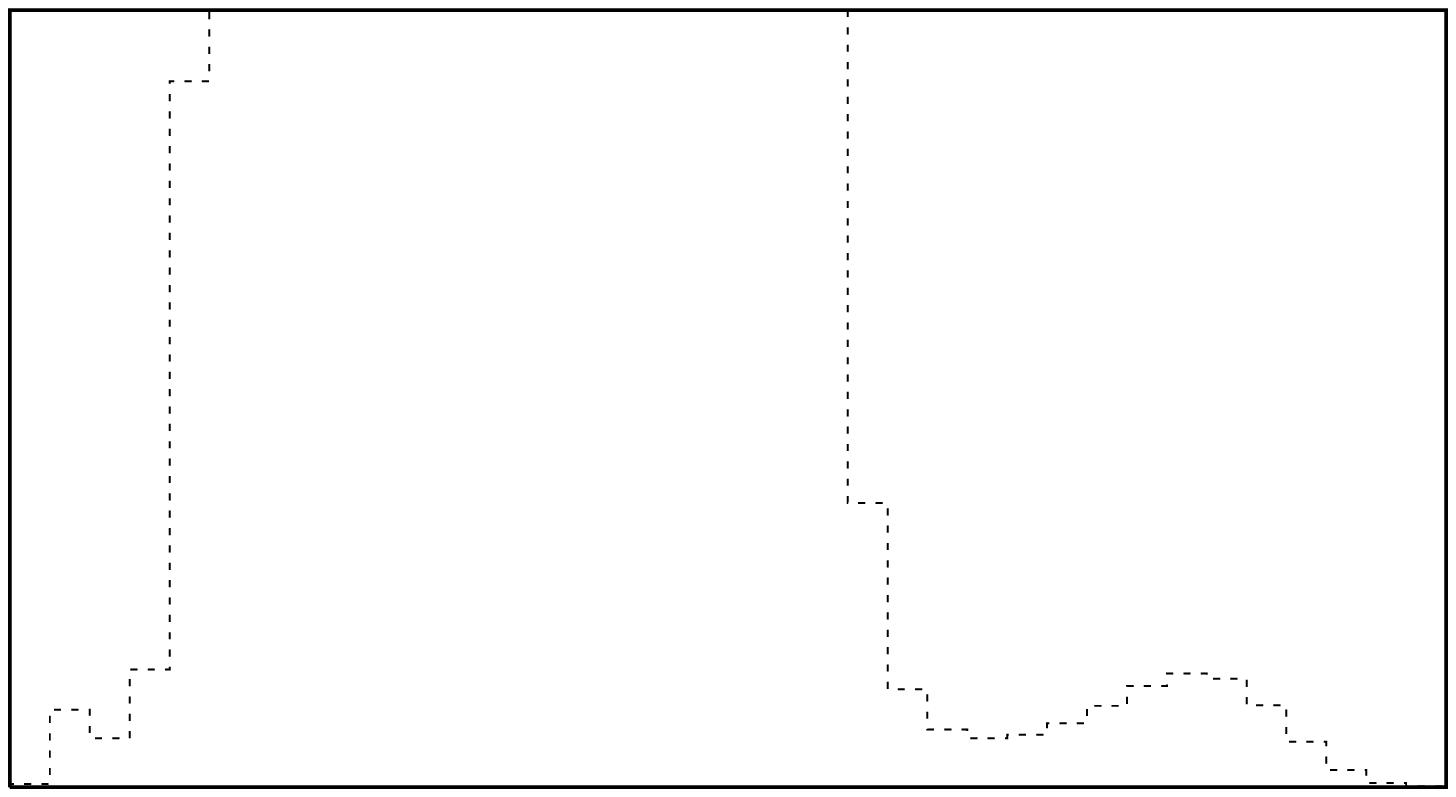}}
\put(197,5){\small$n$}
\put(5,125){\rotl{\texta}}
\put(300,65){\small$\times100$}
\end{picture}
\caption[Histogram for $r/A=36$, $\varphi=120^{\circ}$,
$\nu=0.15$.]{\label{fourmax}
Histogram for a steady state with $\theta=4\pi$ (actually: $r/A=36$,
$\nu=0.15$, $\varphi=2\pi/3$). The dashed line shows $100\rho(n)$.
Consistent with Fig.~\ref{maximin}, we note four maxima: $\rho(1)=0.00020$,
$\rho(7)=0.11316$, $\rho(17)=0.19397$, and $\rho(29)=0.00029$. \hrulefill}
\end{figure}
The threshold concept has its obvious limitations when the photon number
distribution may have more than one peak, as realized in the $\theta=2.09$
case of Fig.~\ref{histos}. More interesting is the question of how many
minima and maxima are there and at which $n$ values. The answer is obtained
after first recognizing this consequence of the recurrence relation
(\ref{2terms}):
\begin{eqns}{rccl}\label{monoton}
\mbox{if}&\mbox{\large$\frac{r}{A}$}\sin^2(\varphi\sqrt{n})>n&
\mbox{then}&\rho(n)>\rho(n-1)\,,\\
\mbox{if}&\mbox{\large$\frac{r}{A}$}\sin^2(\varphi\sqrt{n})<n&
\mbox{then}&\rho(n)<\rho(n-1)\,.
\end{eqns}%
In this way the $n$ ranges in which $\rho(n)$ decreases or increases
monotonically are determined. At the switch-over values we find the maxima
and minima in the photon number distribution. The conditions (\ref{monoton})
acquire a more universal appearance when we introduce a new variable
$\vartheta\equiv\varphi\sqrt{n}$. Then we have
\begin{eqns}{rl}\label{regions}
\sin^2(\vartheta)>(\vartheta/\theta)^2&\mbox{between minima and maxima,}\\
\sin^2(\vartheta)<(\vartheta/\theta)^2&\mbox{between maxima and minima,}
\end{eqns}%
where the pump parameter $\theta=\varphi\sqrt{r/A}$ appears naturally. This
is another reason for using $\theta$ rather than $\varphi$. In passing we
mention that the critical $\vartheta$ values, for which
$\sin^2(\vartheta)=(\vartheta/\theta)^2$
holds, play a great role in the semiclassical theory of the micromaser. In
the context of the quantum treatment discussed here, they determine the $n$
values of the extrema in the photon number distribution. This is illustrated
in Fig.~\ref{maximin}.
The corresponding histogram of $\rho(n)$ should show four maxima, and in
Fig.~\ref{fourmax} it does indeed. 
Of course, not all maxima have to be as pronounced as the two in the
$\theta=2.09\pi$ case of Fig.~\ref{histos}\,(d). Again we note that, as in
(\ref{bel-at-abo}), the thermal photon number $\nu$ is irrelevant in
(\ref{regions}). The number of the extrema in the histogram of the photon
number distribution as well as their locations are solely determined by the
value of $\theta$. Naturally, further details may depend strongly on the
value of $\nu$. 

In the discussion of the plots in Fig.~\ref{OAM-ss-200-0.15} we had remarked
that the pattern roughly repeats itself with a $\theta$ period of $2\pi$.
That may be true for the parameter range of Fig.~\ref{OAM-ss-200-0.15}, but
for other values of the effective rate $r/A$, other thermal photon numbers
$\nu$, or other $\theta$ ranges matters tend to be different. As an example
we take a look at Fig.~\ref{OAM-ss-6-2} where the rate is much smaller,
namely $r/A=6$, and the temperature much higher, $\nu=2$. Rather than the
regular pattern of Fig.~\ref{OAM-ss-200-0.15} we observe an irregular
behavior. The initial oscillation is quickly damped away, but builds up again
at larger values of the Rabi angle $\varphi$, that is: at longer interaction
times. This phenomenon is reminiscent of and related to the so-called
Jaynes-Cummings revivals. Strictly speaking, the latter occur when a
two-level atom is in permanent (resonant) interaction with photons of one
field mode. The Rabi angle is then proportional to the elapsed interaction
time. Although the OAM plots in Fig.~\ref{OAM-ss-6-2} refer to many atoms
with a fixed common interaction time, it is nevertheless clear that the
revivals in this figure are similar --- both in their appearance and in their
physical origin --- to those Jaynes-Cummings revivals.

\savebox{\texta}{\footnotesize $\langle\N\rangle$}
\savebox{\textb}{\footnotesize Fano factor}
\begin{figure}[tbp]
\begin{picture}(390,470)(30,0)
\put(100,15){\epsfysize=220pt\epsffile[85 45 547 411]{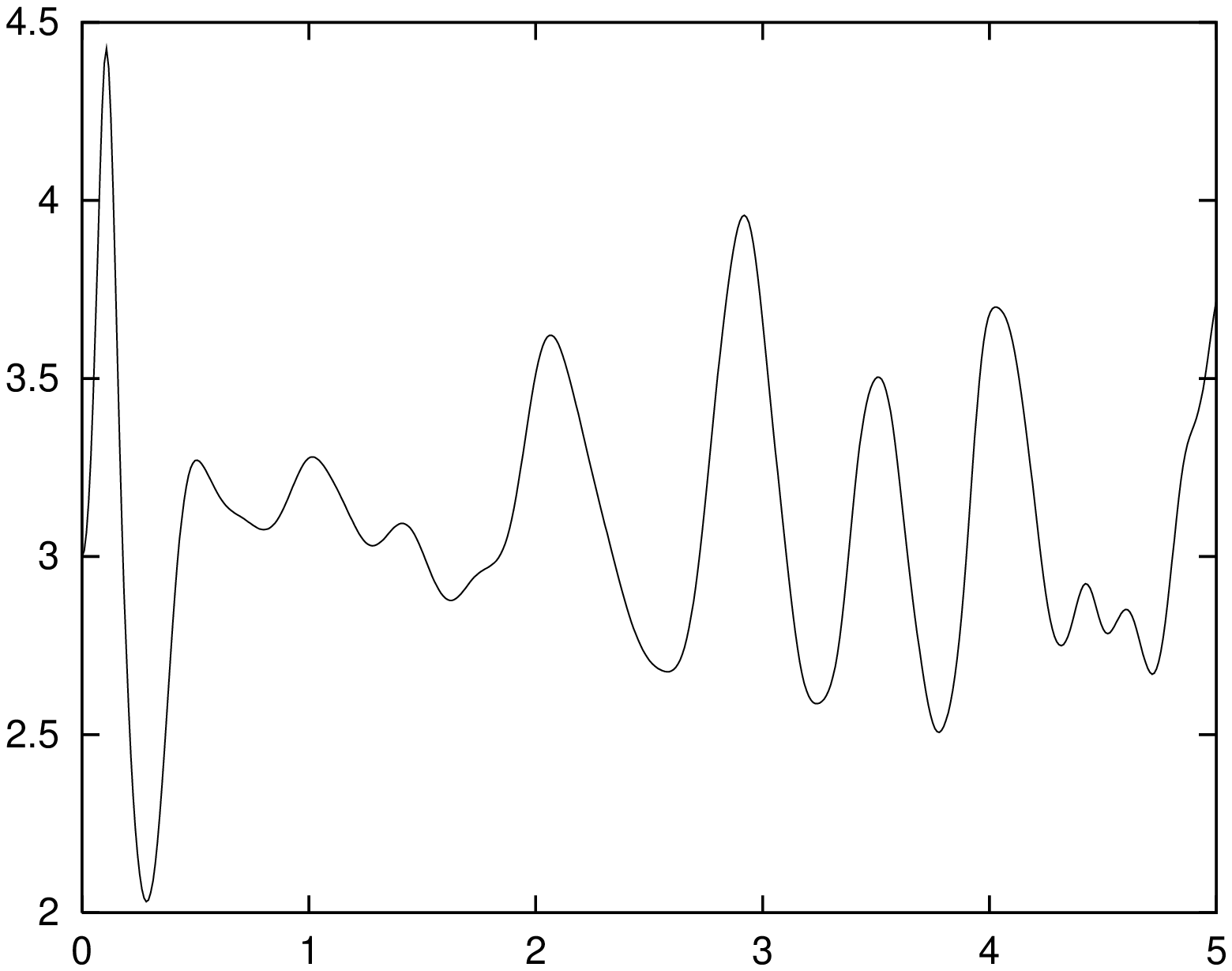}}
\put(100,250){\epsfysize=220pt\epsffile[85 45 547 411]{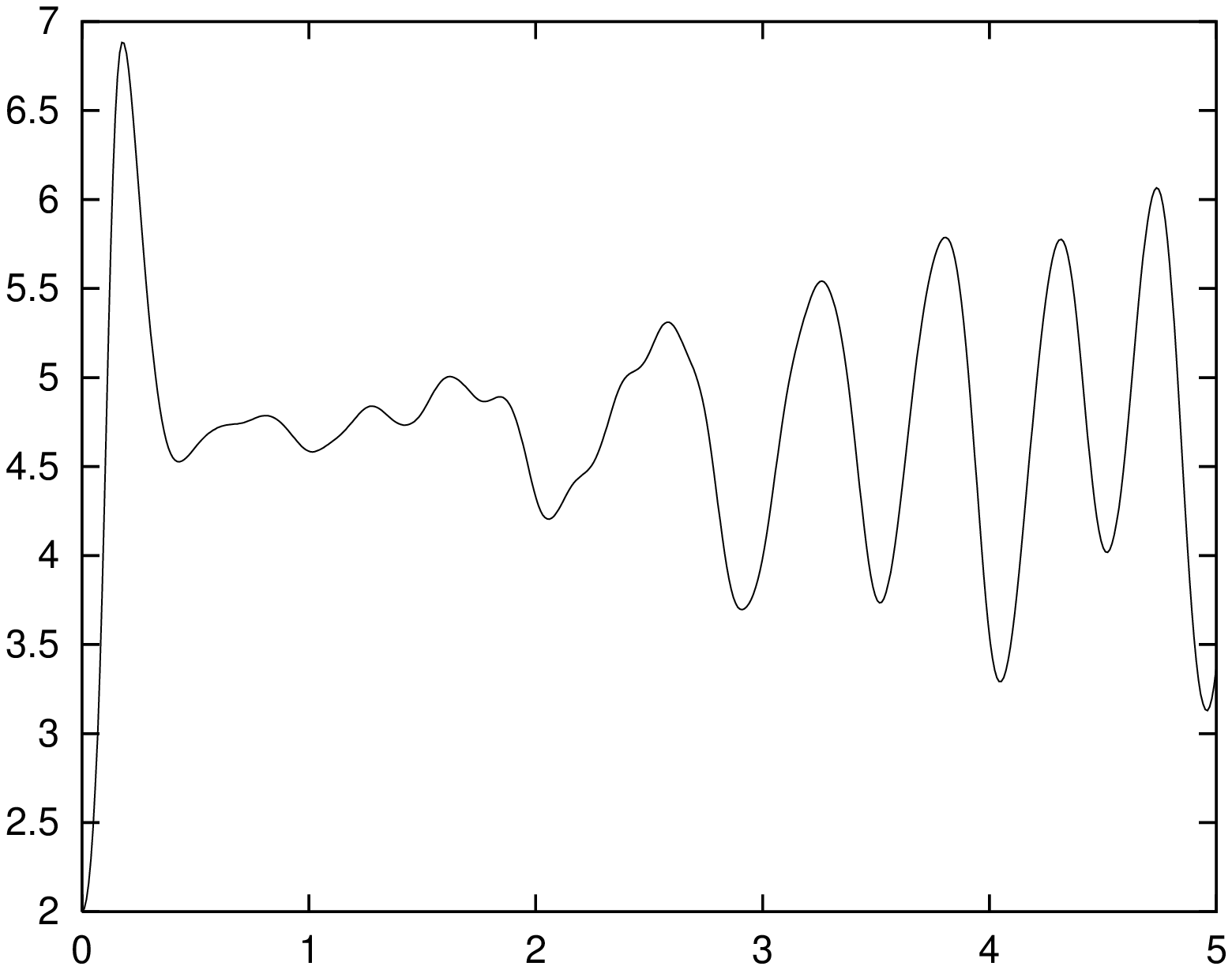}}
\put(234,10){\footnotesize $\varphi/\pi$}
\put(234,245){\footnotesize $\varphi/\pi$}
\put(80,350){\rotl{\texta}}
\put(85,105){\rotl{\textb}}
\end{picture}
\caption[Steady state for $r/A=6,\nu=2$]{\label{OAM-ss-6-2}
Mean photon number and Fano factor in the OAM steady state for an effective
rate of $r/A=6$ and a thermal photon number of $\nu=2$, as a function of the
Rabi angle $\varphi$ in the range from $\varphi=0$ to $\varphi=5\pi$.}
\end{figure}
%
%


\SEC{One-atom maser: Steady state. Energy balance}

One cannot measure the photon state itself in micromaser experiments --- by
counting photons, say --- because that would violate the requirement of a
very large quality factor of the resonator. The experimenter can only look at
the atoms emerging from the resonator after having interacted with the photon
field. The experimental data consists of the statistics of the final atomic
states. For instance, one could count which fraction ends up in the excited
$\UP$ state and which fraction emits a photon and comes out in the deexcited
$\DN$ state. It is an essential task of OAM theory to compute these
fractions, and many other statistical properties of the emerging atoms, and
to relate them, whenever possible, to the properties of the photon state
inside the resonator.

The calculation of the probability for a $\UP\to\DN$ transition during the
passage of the atom through the resonator is based on the result
(\ref{transprob}). We recall that $\ket{\UP,n}\to\ket{\DN,n+1}$ happens with
probability $\sin^2(\varphi\sqrt{n+1})$. Therefore, if the atom encounters a
state
\begin{equation}
\rho(\N)=\sum_{n=0}^{\infty}\ket{n}\rho(n)\bra{n}\,,
\end{equation}
then the total transition probability is given by
\begin{eqns}{rcl}\label{uptodown}
\mbox{prob}(\UP\to\DN)&=&
\sum\limits_{n=0}^{\infty}\rho(n)\sin^2(\varphi\sqrt{n+1})\\
&=&\mbox{tr}\{\rho\sin^2[\varphi(aa^{\dag})^{1/2}]\}\\
&=&\langle \sin^2[\varphi(aa^{\dag})^{1/2}]\rangle\,.
\end{eqns}%
This probability can be computed immediately once the steady state $\rho$ has
been determined in accordance with (\ref{OAMss}). In a series of challenging
experiments, actual measurements have been performed (see~\cite{rev:MPQ} and
the references therein, in particular~\cite{GR+HW+NK87} and \cite{GR+HW90}),
and very good agreement with the theoretical predictions has been found.

When we multiply the transition probability (\ref{uptodown}) by the beam rate
$r$ and the energy per photon $\hbar\omega$, we obtain the rate at which the
photon field gains energy: $r\,\hbar\omega\,\mbox{prob}(\UP\to\DN)$. In view
of Eq.~(\ref{numbdecay}), the energy loss rate is likewise given by the
product of the decay rate $A$, the energy per photon $\hbar\omega$, and the
difference between the actual mean photon number $\langle\N\rangle$ and the
thermal photon number $\nu$: $A\,\hbar\omega\,(\langle\N\rangle-\nu)$. In
steady state these two rates must be equal. After canceling the common
factor of $\hbar\omega$ we thus arrive at a fifth central result of
micromaser theory:
\begin{equation}\label{energybal}
\fbox{\parbox{250pt}{
\begin{center}
\parbox{220pt}{In the steady state of a one-atom maser, pumped by resonant
$\UP$ atoms, the energy balance implies the equality}
\begin{displaymath}
\underbrace{r\langle \sin^2[\varphi(aa^{\dag})^{1/2}]\rangle}_{\mbox{
\small$\propto$ energy gain rate}}=
\underbrace{\ \ A(\langle\N\rangle-\nu)\ \ }_{\mbox{\small$\propto$ energy
loss rate}}.
\end{displaymath}
\parbox{220pt}{The expectation value on the left-hand side can be measured
experimentally.}
\end{center}
}}
\end{equation}
We emphasize that by determining the fraction of atoms that have emitted a
photon one can experimentally measure the mean photon number inside the
resonator. The link between the two quantities is provided by the theory of
the OAM and stated in (\ref{energybal}). 

The way in which we derived this statement simply appealed to the physical
significance of the expressions on the two sides. A more formal derivation
employs the two-term recurrence relation (\ref{2terms}) in the form
\begin{equation}
r\sin^2[\varphi(aa^{\dag})^{1/2}]\rho(\N)=A\big[(\nu+1)\rho(\N+1)-\nu\rho(\N)
\big](\N+1)\,.
\end{equation}
Upon taking the trace on both sides, we find
\begin{equation}
r\langle\sin^2[\varphi(aa^{\dag})^{1/2}]\rangle=
A(\nu+1)\langle\N\rangle-A\nu\langle\N+1\rangle\,,
\end{equation}
which, after an elementary simplification, is recognized to be identical to
the statement in (\ref{energybal}).


\SEC{One-atom maser: Trapped states}

\savebox{\texta}{\small$\rho(n)$}
\savebox{\textb}{\small$\rho(n)$}
\begin{figure}[tb]
\begin{picture}(390,230)(0,0)
\put(15,15){\epsfxsize=170pt\epsffile[99 50 386 402]{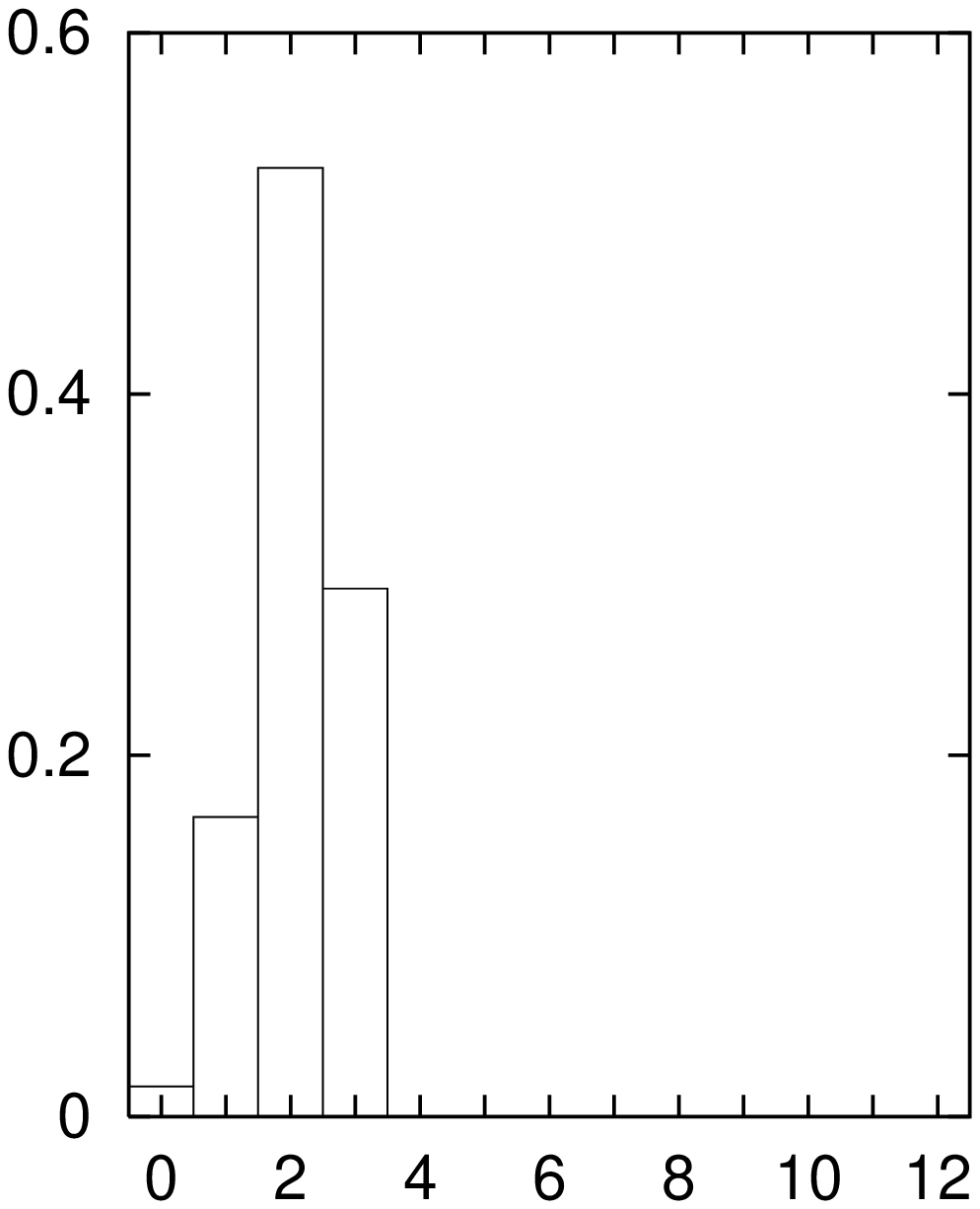}}
\put(220,15){\epsfxsize=170pt\epsffile[99 50 386 402]{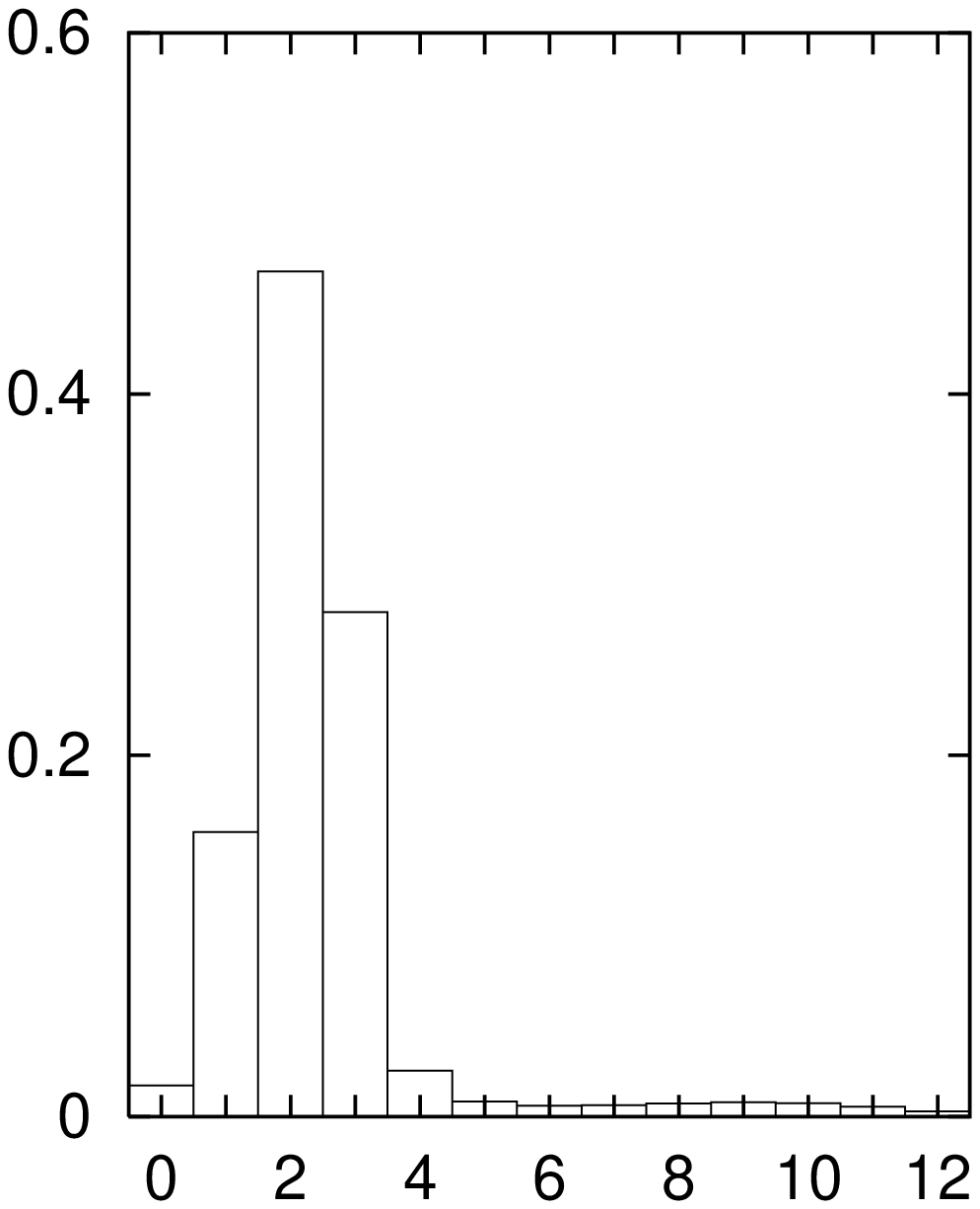}}
\put(5,118){\rotl{\texta}}
\put(210,118){\rotl{\textb}}
\put(105,5){\small$n$}
\put(310,5){\small$n$}
\put(155,195){\small(a)}
\put(360,195){\small(b)}
\end{picture}
\caption[Histograms: trapped/not trapped]{\label{traphistos}
Histograms for the OAM steady states obtained for $r/A=10$, $\varphi=\pi/2$,
and (a) $\nu=0$, (b) $\nu=0.1$. In case (a) the trapping limits the possible
photon numbers to the values $n=0,1,2,3$. No trapping occurs in case (b).
\hrulefill}
\end{figure}

At zero temperature there are no thermal photons: $\nu=0$. Then the
contributions [{\sc c}] and [{\sc e}] are not present in Eqs.~(\ref{diagME})
and (\ref{3terms}) and in Figs.~\ref{gainloss} and \ref{detbal}. As a
consequence, the steady state (\ref{OAMss}) of the one-atom maser is
particularly simple, viz.
\begin{equation}\label{OAMss0}
\left[\ \rho^{\stext{(SS)}}(\N)\ \right]=\rho(\N)
=\rho(0)\,\frac{(r/A)^{\disp\N}}{(\N)!}\,\prod_{n=1}^{\disp\N}
\sin^2(\varphi\sqrt{n})\,.
\end{equation}
In contrast to the general $\nu\neq0$ case of (\ref{OAMss}), here the
individual factors of the product might vanish. Thus we note this property of
(\ref{OAMss0}):
\begin{eqns}{rccl}\label{trapping}
\mbox{if} & \sin^2(\varphi\sqrt{N})=0 & \mbox{holds for} &
N=1,2,3,\ldots\,,\\
\mbox{then} & \rho(\N)=0 & \mbox{holds for} & \N\geq N\,.
\end{eqns}%
One says that the photon number is `trapped' below $\N=N$.

Such trapped states are very characteristic of the OAM dynamics. For purely
energetic reasons, the mean photon number $\langle\N\rangle$ could be as
large as the effective pump rate $r/A$ [recall Eq.~(\ref{energybal}), here
for $\nu=0$]. The specific OAM dynamics, however, may limit the photon number
to much smaller values. 
We illustrate this in Fig.~\ref{traphistos}, where the histogram for
$\varphi=\pi/2$, that is: $N=4$ in (\ref{trapping}), shows nonzero
probabilities only for the photon numbers $n=0,1,2,3$, although the effective
rate, $r/A=10$, is more than twice as large. A second plot in this figure
demonstrates how a small number of thermal photons ($\nu=0.1$) removes the
trapping.

\savebox{\texta}{\small$\langle\N\rangle$}
\begin{figure}[tb]
\begin{picture}(390,220)(0,0)
\put(20,15){\epsfxsize=360pt\epsffile[99 50 540 205]{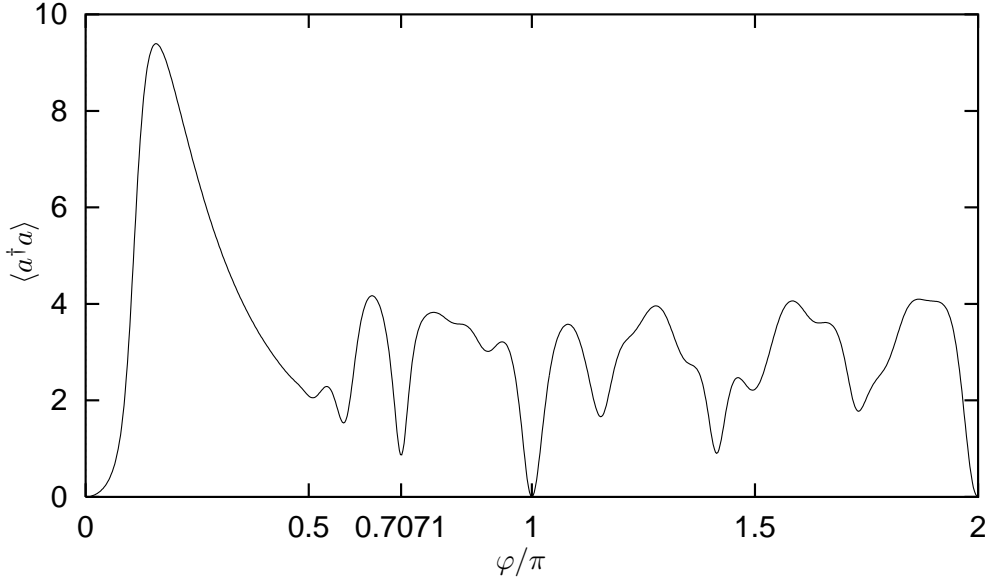}}
\put(5,113){\rotl{\texta}}
\put(190,5){\small$\varphi/\pi$}
\end{picture}
\caption[OAM photon number, $r/A=10,\nu=0$]{\label{OAM-ss-10-0}
Mean photon number in the OAM steady states obtained for an effective pump
rate of $r/A=10$ at zero temperature ($\nu=0$), as a function of the Rabi
angle $\varphi$ in the range $0\leq\varphi\leq2\pi$. Please note the dips at
$\varphi=\pi$ and  $\varphi=2\pi$ that indicate trapped vacuum states.
\hrulefill}
\end{figure}

A very remarkable trapped state is the trapped vacuum state, which is
realized if $\varphi$ equals an integer multiple of $\pi$. In
Fig.~\ref{OAM-ss-10-0}, the trapped vacuum states at $\varphi=\pi$ and
$\varphi=2\pi$ are visible as pronounced dips in the plot of the mean photon
number. Note also the minimum at $\varphi=\pi/\sqrt{2}$ that corresponds to a
trapping below $\N=2$.

When the vacuum is trapped, the atoms perform one or more complete Rabi
cycles while traversing the resonator. In other words, each atom first emits
a photon and then reabsorbs it --- perhaps repeatedly --- eventually leaving
the resonator in the excited state $\UP$ in which it entered. This is, of
course, the $n=0$ situation of Eq.~(\ref{emitnot}). Under these circumstances
it does not matter how frequent the one-atom events are, nor if they occur in
accordance with the Poisson statistics of (\ref{waiting}). It is simply
impossible to get photons into the resonator.

\SEC{Approximations in one-atom-maser theory}

A direct experimental demonstration of the trapped vacuum state has not been
achieved as yet. To understand why, we recall that the theoretical result
(\ref{OAMss0}) is based upon various approximations. Let us reconsider them
and their significance for the trapped vacuum state. In this context, the
most important approximations are:
\begin{list}{}{\setlength{\leftmargin}{10pt} }
\item\hspace*{-13pt} \underline{The interaction time is assumed to be the
same for all atoms.}\\[4pt]
This assumption has two aspects:
\begin{list}{}{\setlength{\topsep}{-3pt}
\setlength{\leftmargin}{10pt}\setlength{\itemsep}{0pt}}
\item\hspace*{-10pt}{\em The atoms should have the same velocity
$v$.}\\[2pt]
Differing velocities lead to differing Rabi angle $\varphi$. Fortunately,
however, the trapping condition requires $\varphi=\pi$, where
$\sin^2(\varphi)$ has a minimum. Therefore, deviations from the ideal
$\varphi$ value of $\pi$ do not change anything to first order. A velocity
control to about one percent is good enough, and that can be realized
experimentally.
\item\hspace*{-10pt}{\em Atomic decays to other levels do not
occur.}\\[2pt]
The Rydberg states of Fig.~\ref{Rb85} do not meet this condition sufficiently
well; a few percent of the atoms make ultraviolet transitions to low lying
levels and thus stop interacting with the photons in the privileged mode
before they actually leave the resonator. Therefore, one has to use Rydberg
states that live even longer, the so-called circular states in which both the
angular and the magnetic quantum number are maximal~\cite{CIRC}. The masing
of such circular states has already been demonstrated
experimentally~\cite{CIRCMASER}. A one-atom-maser operation is likely to be
realized in the near future.
\end{list}
\item\hspace*{-9pt}\underline{The transition frequency $\Omega$ of
Eq.\hspace{5.075pt}(\ref{JCMham}) is assumed to be equal to the photon}\break
\hspace*{-9pt}\underline{frequency $\omega$ for all atoms.}\\[4pt]
This has also two main aspects:
\begin{list}{}{\setlength{\topsep}{-3pt}
\setlength{\leftmargin}{10pt}\setlength{\itemsep}{0pt}}
\item\hspace*{-10pt}{\em The resonator must be mechanically stable.}\\[2pt]
Apparently one does not have to worry much about this. I am told that the
detuning resulting from mechanical instabilities is less than
100\,Hz.\footnote{A detuning should not be compared with the transition
frequency of $\sim21.5$\,GHz, but rather with the Rabi frequency of
$\sim14$\,kHz.}
\item\hspace*{-10pt}{\em Electric stray fields must be small.}\\[2pt]
Here the concern is about Stark shifts of the atomic levels. Inasmuch as the
stray fields are prominent only in the entry and exit ports of the resonator,
and not so much inside where the relevant atom-photon interaction takes
place, they are not very critical. In other experiments, however, in which
the atoms are prepared in coherent superpositions of $\UP$ and $\DN$, stray
fields are a great nuisance because the differential Stark shift changes the
phase relations between $\UP$ and $\DN$ and thus may destroy the desired
superposition.
\end{list}
\item\hspace*{-13pt} \underline{The photon damping is neglected during the
atom-photon interaction.}\\[4pt]
This is harmless. Indeed, the Jaynes-Cummings model can be extended by the
inclusion of photon damping. One then finds~\cite{DB} that the trapping
condition of (\ref{trapping}) is altered slightly, but trapping still
exists.
\item\hspace*{-13pt} \underline{Collective events are not taken into
account.}\\[4pt]
This is a critical point. We shall deal with it in some detail.
\end{list}

\SEC{Two-atom events}

A typical number for $AL/v$ in the Garching micromaser experiments is
$10^{-4}\pi$. For an effective beam rate of $r/A=10$, the probability
(\ref{OAprob}) for a one-atom event is then
$\exp(-2rL/v)=\exp[-2(r/A)\times(AL/v)]=99.4\%$. In this situation, only
$0.6\%$ of the atoms participate in collective events, and most of those in
two-atom events.

\savebox{\texta}{\small$\rho(n)$}
\begin{figure}[tb]
\begin{picture}(390,240)(0,0)
\put(20.5,117){\epsfxsize=138pt\epsffile[99 48 391 445]{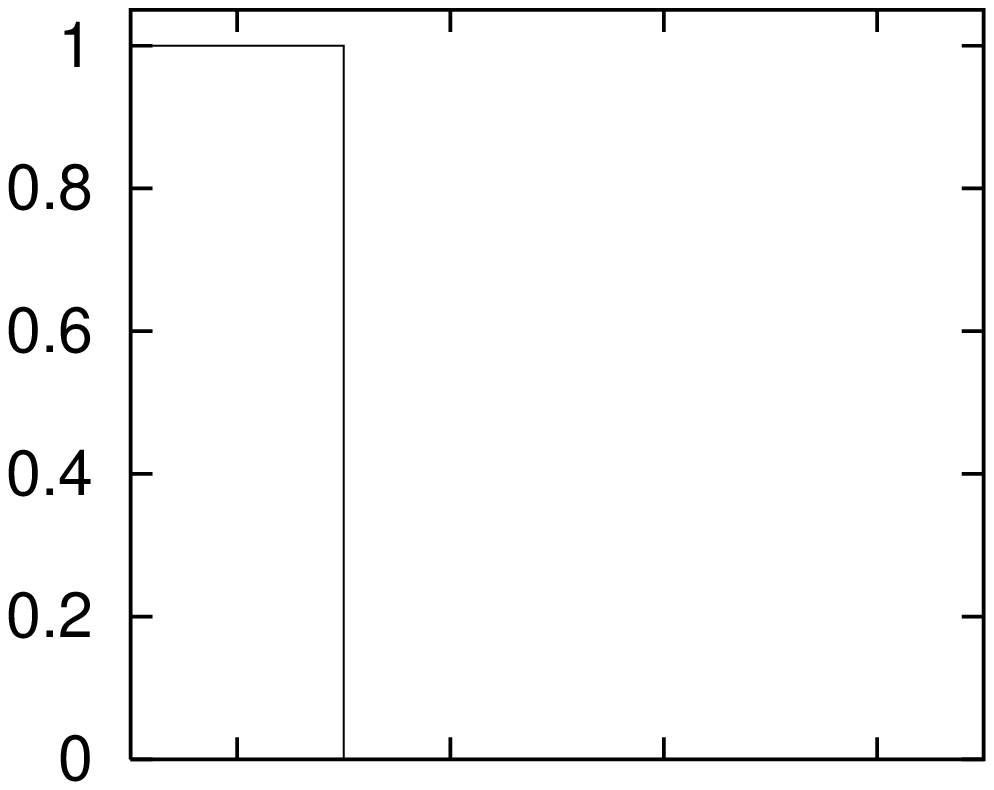}}
\put(136.5,117){\epsfxsize=138pt\epsffile[99 48 391 445]{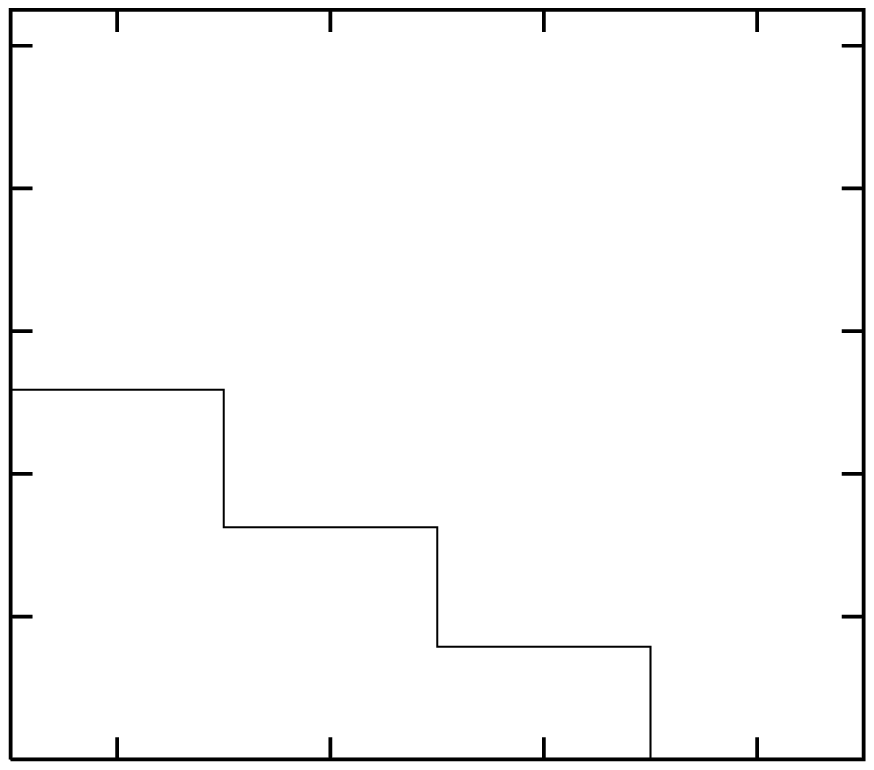}}
\put(253,117){\epsfxsize=138pt\epsffile[99 48 391 445]{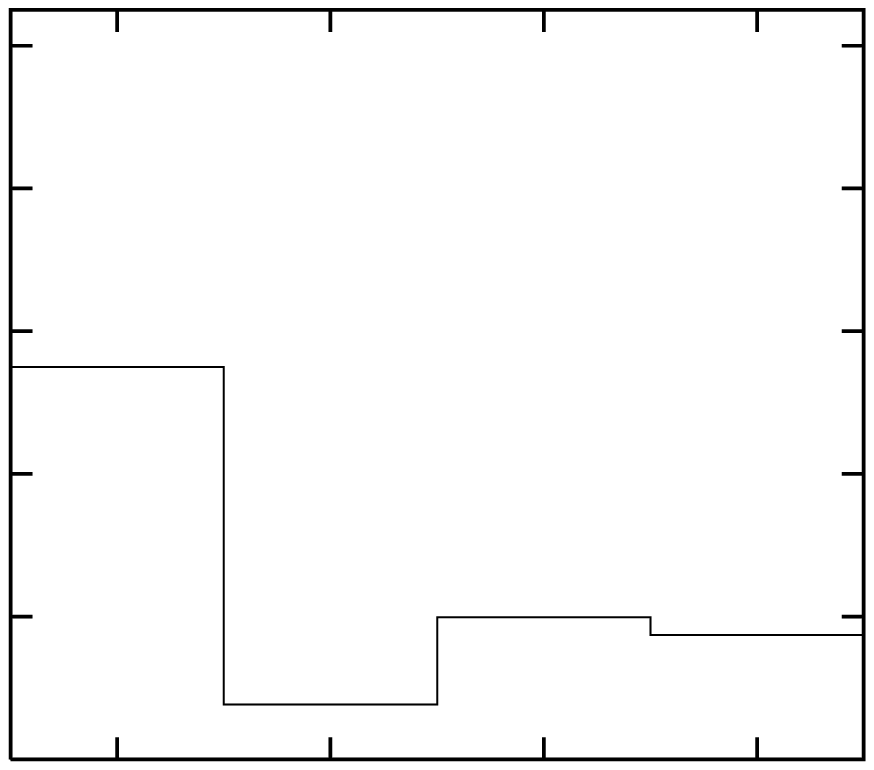}}
\put(20.5,15){\epsfxsize=138pt\epsffile[99 48 391 445]{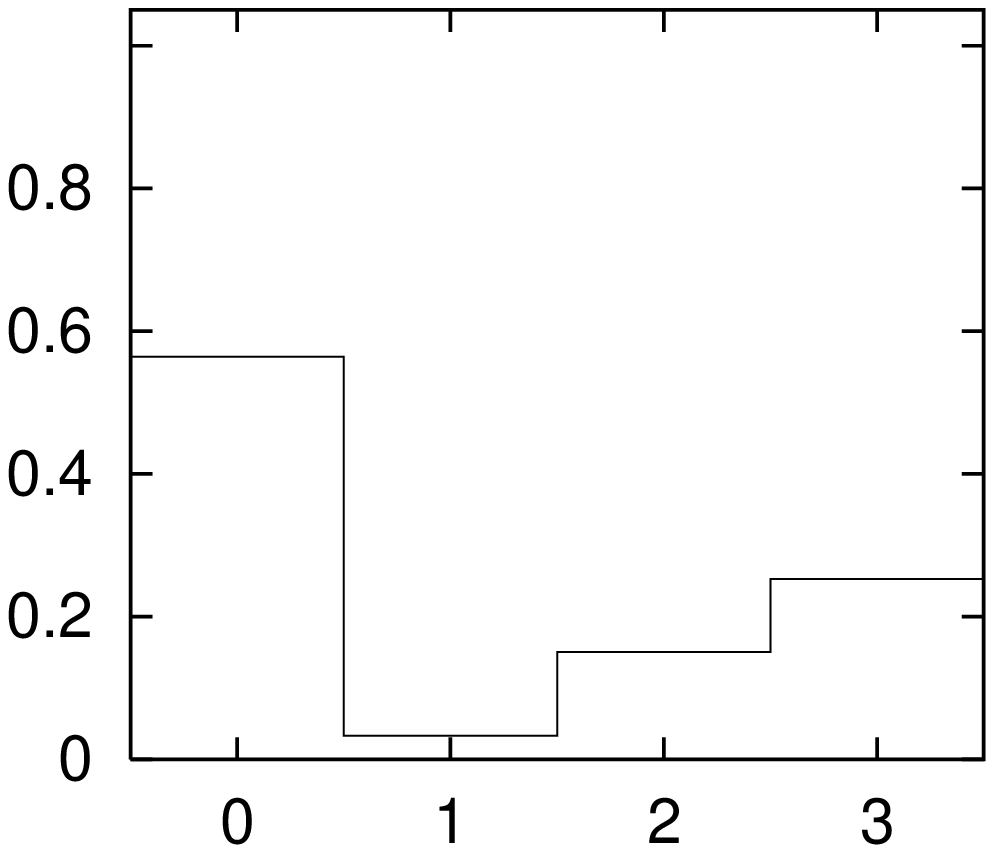}}
\put(136.5,15){\epsfxsize=138pt\epsffile[99 48 391 445]{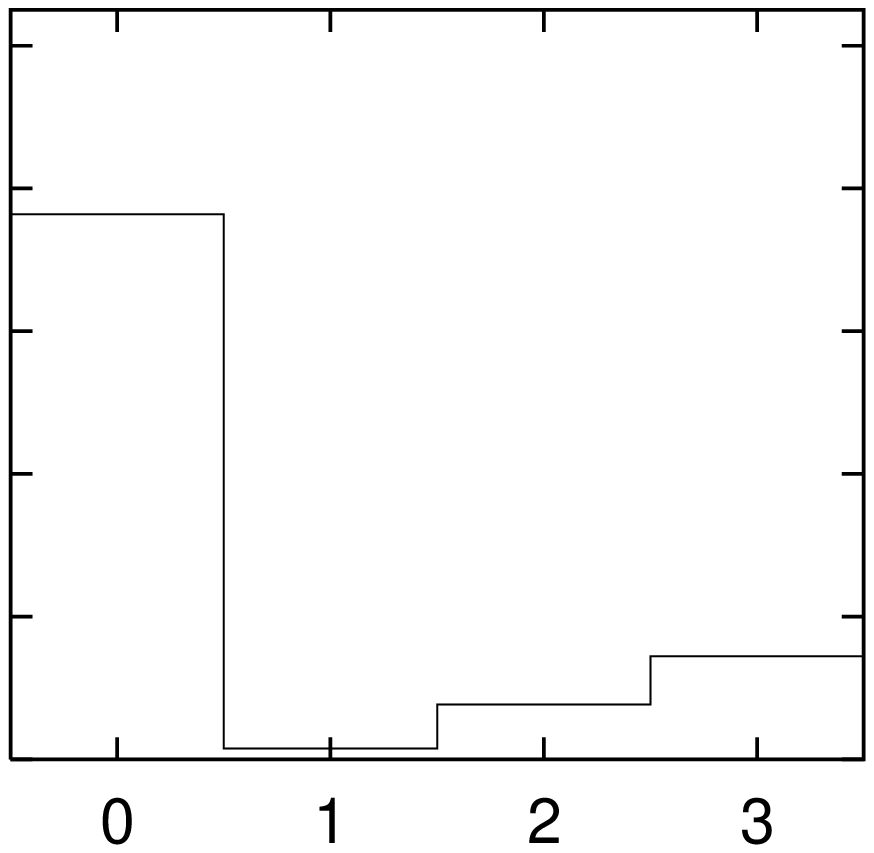}}
\put(253,15){\epsfxsize=138pt\epsffile[99 48 391 445]{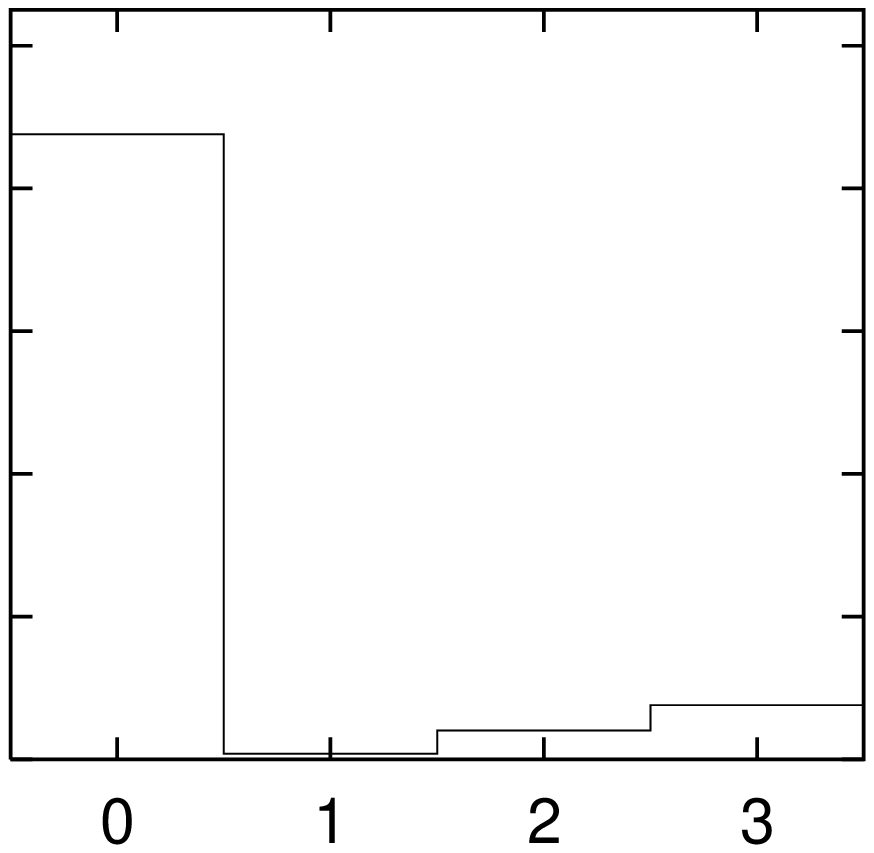}}
\put(5,124){\rotl{\texta}}
\put(212,5){\small$n$}
\put(130,212){\small (a)}
\put(246.5,212){\small (b)}
\put(363,212){\small (c)}
\put(130,110){\small (d)}
\put(246.5,110){\small (e)}
\put(363,110){\small (f)}
\end{picture}
\caption[Histograms, damage after a pair]{\label{pairhistos}
Histograms of (a) the trapped vacuum state; (b) the state immediately after
the passage of an atom pair; and after subsequent one-atom operation for a
period during which (c) two, (d) five, (e) one hundred, or (f) two hundred
one-atom events occur on average. \hrulefill}
\end{figure}

Although these collective events are very rare, they may cause dramatic
effects nevertheless. In particular, they destroy the trapped vacuum state
completely. The physical reason is the following. After two atoms, say, have
been in the resonator simultaneously, the photon field is no longer in the
vacuum state, there are nonzero probabilities for one or two photons. Single
atoms that follow will then emit more photons and so drive the maser field
further away from the vacuum. This is demonstrated in Fig.~\ref{pairhistos}.
Eventually, after a large number of one-atom events, the maser field will
return to the vacuum state. The analysis given below shows that {\em very
many} one-atom events are needed to undo the damage caused by a single
two-atom event, so many indeed that another two-atom event is very likely to
occur in the meantime. Therefore, we have to conclude that, despite their
rare occurrence, the collective events prevent the observation of the
trapped vacuum state in a micromaser with a poissonian pump beam.

Consider the OAM dynamics of Eq.~(\ref{diagME}) at zero temperature ($\nu=0$)
and for $\varphi=\pi$,
\begin{nareqns}{rl}\label{diagME0}
\mbox{\Large$\partt$}\rho_t(\N)=&
r_1\sin^2\left[\pi(\N)^{1/2}\right]\rho_t(\N-1)
+A(\N+1)\rho_t(\N+1)\\
&-\left\{r_1\sin^2\left[\pi(\N+1)^{1/2}\right]
+A\N\right\}\rho_t(\N)\,,
\end{nareqns}%
where $r_1$ is the rate of one-atom events; its relation to the beam rate
$r$ is given by $r_1=r\,\exp(-2rL/v)$. One verifies easily that the vacuum
state $\rho_{\stext{vac}}=\delta(\N,0)$ is the steady state of
(\ref{diagME0}).

For $\varphi=\pi$, the trapping condition is not only obeyed by $N=1$, but
also by $N=4,9,16,\ldots$, so that the threshold from three to four photons
cannot be crossed by one-atom events. Now, after a two-atom event has
happened to an initial vacuum state, the probabilities for one or two photons
will be nonzero, but those for three or more photons still vanish. As long as
only one-atom events occur thereafter, the probabilities for more than three
photons remain zero. Thus a fitting ansatz for $\rho_t(\N)$ is
\begin{equation}\label{pairsrho}
\rho_t=
\alpha^{(0)}_t\delta(a^{\dag}a,0)+\alpha^{(1)}_t\delta(a^{\dag}a,1)
+\alpha^{(2)}_t\delta(a^{\dag}a,2)+\alpha^{(3)}_t\delta(a^{\dag}a,3)\,,
\end{equation}
wherein the time-dependent numerical coefficients $\alpha^{(0,1,2,3)}_t$ are
the respective probabilities to have no, one, two, or three photons in the
resonator. The sum of these probabilities is 100\% --- this is the
normalization of $\rho$ to unit trace --- and that requires
$\alpha^{(0)}=1-\alpha^{(1)}-\alpha^{(2)}-\alpha^{(3)}$. 

At $t=0$, we have $\alpha^{(3)}_0=0$, of course, and typical numbers for the
initial values of $\alpha^{(1)}_t$ and $\alpha^{(2)}_t$ are~\cite{pairs}
\begin{equation}
\alpha^{(1)}_0=\frac{1}{3}\left[\sin(\pi\sqrt{6})\right]^2=0.3250\,,\quad
\alpha^{(2)}_0=\frac{8}{9}\left[\sin(\pi\sqrt{3/2})\right]^4=0.1575\,.
\end{equation}
As implied by (\ref{diagME0}), the time dependence of these coefficients
is determined by
\begin{equation}\label{33matrix}
\renewcommand{\arraystretch}{1.6}
\left[
\frac{1}{A}\frac{\diff{}}{\diff{t}}
+\left(
\begin{array}{ccc}
1+\Rone\sin^2(\pi\sqrt{2}) & -2 & 0 \\
-\Rone\sin^2(\pi\sqrt{2}) & 2+\Rone\sin^2(\pi\sqrt{3}) & -3 \\
0 & -\Rone\sin^2(\pi\sqrt{3}) & 3 \\
\end{array}
\right) \right]
\left(
\begin{array}{c}
\alpha^{(1)}_t \\ \alpha^{(2)}_t \\ \alpha^{(3)}_t
\end{array}
\right)=0\,.
\end{equation}
The time scale for the return to the vacuum is set by the eigenvalues of this
$3\times3$ matrix, more precisely: by the smallest one. For $r/A=10$,
$r_1/A=9.937$ these eigenvalues are
\begin{equation}
13.974\,,\quad 6.7225\,,\quad 
0.063869=\frac{r_1/A}{155.6}\,.
\end{equation}
Please note that the smallest one is {\em very small\/} indeed. It takes
hundreds of one-atom events to get close to the vacuum state; see
Fig.~\ref{pairhistos} again, where the photon states are reported for
$r_1t=0,2,5,100,200$. During such a long time, another collective event is
very likely to occur, so that the vacuum state will not be reached at all. 

\savebox{\texta}{\small prob($\UP\to\DN$)}
\begin{figure}[tbp]
\begin{picture}(390,290)(-8,0)
\put(20,15){\epsfxsize=360pt\epsffile[85 48 553 408]{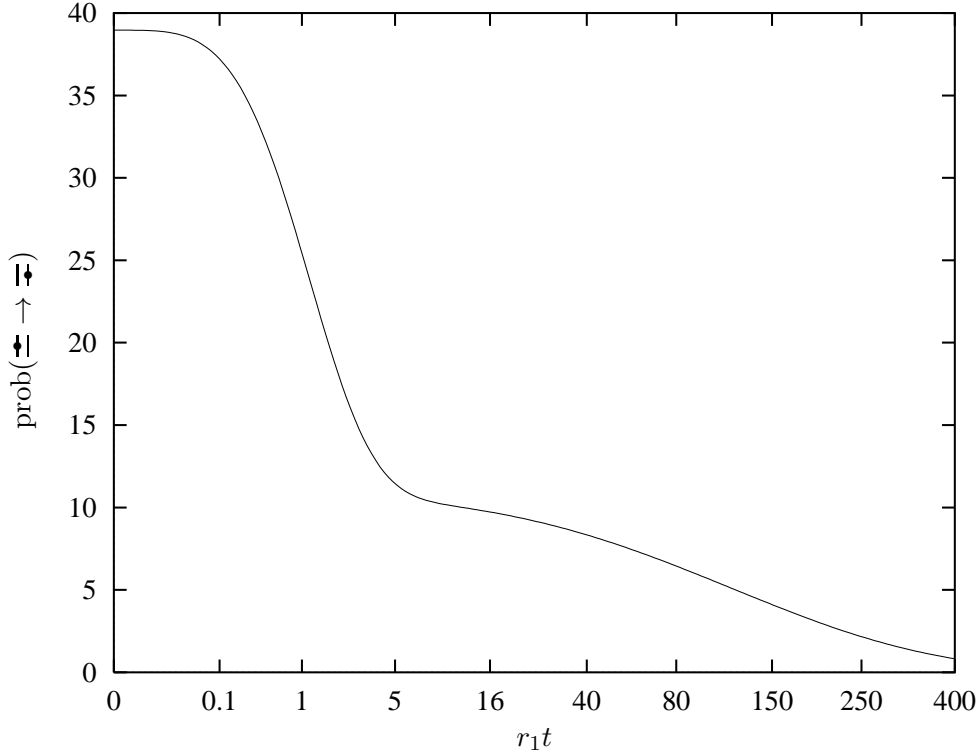}}
\put(197,5){\small$r_1t$}
\put(5,125){\rotl{\texta}}
\end{picture}
\caption[Emission probability after passage of a pair]{\label{pairemit}
Time dependence of the emission probability of atoms in one-atom events
after the passage of an atom pairs, for $0\leq t\leq400/r_1$. The abscissa is
linear in $(r_1t)^{1/4}$. \hrulefill}
\end{figure}

To be more specific, we note that the experimental signature for the trapped
vacuum state is the observation of all atoms emerging in the excited $\UP$
state. Therefore, an appropriate numerical criterion should involve the
probability that the atom emits a photon if it traverses the resonator at
time~$t$. This probability is given in Eq.~(\ref{uptodown}). For states of
the form (\ref{pairsrho}) it reads
\begin{equation}
\mbox{prob}_t(\UP\to\DN)=\alpha_t^{\stext{(1)}}\sin^2(\pi\sqrt{2})
+\alpha_t^{\stext{(2)}}\sin^2(\pi\sqrt{3})\,.
\end{equation}
This probability is plotted in Fig.~\ref{pairemit}. If we now require that
99\% of the atoms emerge in the upper state $\UP$, then the emission
probability must get as small as 1\% before we would say that the vacuum
state has been reached again. The time $t=370/r_1$ has to go by for the
parameters we have been using consistently, and the probability that at least
one more collective event takes place in the meantime is 68.8\%~\cite{pairs}.
This number is discouragingly large --- it is clear that one needs to operate
the micromaser with a pump beam that has considerably fewer collective
events, if the trapped vacuum state is to be observed.

The small eigenvalue of the $3\times3$ matrix in (\ref{33matrix}) is one
theoretical indication that very long time scales are present in the dynamics
of the micromaser. In this context, it is worth mentioning that very recent
experiments~\cite{bistable} have demonstrated the presence of extremely large
time scales in rather different dynamical regimes of the micromaser as well.

\SEC{Correlations among emerging atoms}

We have remarked above that the photon state cannot be observed directly in
micromaser experiments. The only reproducible data concerns the emerging
atoms, more specifically: their statistical properties. Experimental tests of
the micromaser theory must therefore compare the measured atom statistics to
the theoretically predicted ones. We have seen one elementary example
already, in the context of Eq.~(\ref{energybal}). On the one hand, one can
calculate the transition probability (\ref{uptodown}); on the other hand, it
can be measured by counting the atoms that end up in the $\UP$ or the $\DN$
state.

More subtle are statistical predictions about correlations among emerging
atoms. The general theory for their calculation is rather
involved~\cite{OAMstatistics} and well beyond the scope of this tutorial
review. We shall therefore be content with presenting two rather simple
examples. They serve the purpose of conveying the spirit in which such
calculations are done without requiring too much technical detail.

Consider a OAM at zero temperature ($\nu=0$) operated with a Rabi angle of
$\varphi=\pi/\sqrt{2}$, so that the trapped state with one photon at most,
\begin{equation}
\rho^{\stext{(SS)}}=\frac{1}{A+r\sin^2(\pi/\sqrt{2})}
\left[A\delta(\N,0)+r\sin^2(\pi/\sqrt{2})\delta(\N,1)\right]\,,
\end{equation}
is the steady state. The meaning of this $\rho^{\stext{(SS)}}$ is this: it is
to be used for statistical predictions, if nothing else is known. For
instance, the (a priori) probability for an atom to end up in the $\DN$
state is
\begin{equation}\label{apriori}
\mbox{tr}\left\{\sin^2[\pi(\N/2)^{1/2}]\rho^{\stext{(SS)}}(\N)\right\}
=\frac{A\sin^2(\pi/\sqrt{2})}{A+r\sin^2(\pi/\sqrt{2})}\,,
\end{equation}
as stated in Eq.~(\ref{uptodown}).

Now suppose that we have just registered, at $t=0$, an atom in the $\DN$
state. What is then the probability that the next atom emerges also in the
$\DN$ state? We are here asking for a conditional probability, rather than
for an a priori probability. The steady state $\rho^{\stext{(SS)}}$ is not
appropriate any more. In its stead we must employ that time dependent state
$\rho_t$ which correctly accounts for the imposed conditions. Then
\begin{equation}
\mbox{prob}_t(\UP\to\DN)=
\mbox{tr}\left\{\sin^2[\pi(\N/2)^{1/2}]\rho_t\right\}
\end{equation}
is the transition probability if the next atom comes after the elapse of
time~$t$. And the probability that the next atom arrives between $t$ and
$t+\diff{t}$ is $\diff{t}\,r\exp(-rt)$, where we remind ourselves of the
Poisson formula (\ref{waiting}). Upon putting things together, we find that
the asked-for probability that the next atom also emits a photon is given by
\begin{equation}
\mbox{prob(next in $\DN$)}=\int_0^{\infty}\diff{t}\,re^{-rt}\,
\mbox{tr}\left\{\sin^2[\pi(\N/2)^{1/2}]\rho_t\right\}\,.
\end{equation}
We still have to say how the correct $\rho_t$ is calculated. This requires to
find (i) the appropriate initial state $\rho_0$ and (ii) the appropriate
master equation for~$\rho_t$. 

At $t=0$, we know that an atom has just been registered in the $\DN$ state.
Prior to its passage through the resonator we had no additional information,
so that then $\rho^{\stext{(SS)}}$ applies. Immediately after the
observation of the initial $\DN$ atom, however, the photon state is reduced
to account for this information,
\begin{equation}\label{reduction}
\mbox{state reduction:}\quad \rho^{\stext{(SS)}}\longrightarrow\delta(\N,1)
\equiv\rho_0\,.
\end{equation}
In other words: after the registration of the atom emerging in the $\DN$
state we know with certainty that there is one photon in the resonator.

Until the next atom arrives, there are no other atoms traversing the
resonator. Consequently, $\rho_t$ is determined by the master equation
(\ref{unpumped}) of the unpumped resonator (here for $\nu=0$). So we find
first
\begin{equation}
\rho_t=\left(1-e^{-At}\right)\delta(\N,0)+e^{-At}\delta(\N,1)\,,
\end{equation}
then
\begin{equation}\label{condoft}
\mbox{prob}_t(\UP\to\DN)=\left(1-e^{-At}\right)\sin^2(\pi/\sqrt{2})\,,
\end{equation}
and finally
\begin{equation}
\mbox{prob(next in $\DN$)}=\left(1-\frac{r}{A}\right)\sin^2(\pi/\sqrt{2})\,.
\end{equation}
This answers the question asked above. Compared with the a priori transition
probability in (\ref{apriori}), this conditional transition probability is
smaller.

The time dependent transition probability (\ref{condoft}) vanishes for $t=0$
and approaches $\sin^2(\pi/\sqrt{2})$ for $t\to\infty$. Both properties are
easily understood on physical grounds. At $t=0$, that is: immediately after
the first atom was found in the $\DN$ state, the trapping condition forbids
the emission of another photon into the resonator. And if no atom has
traversed the resonator for a time long compared to the photon life time
$1/A$ --- this is the physical significance of the mathematical $t\to\infty$
limit --- the photon field is in the vacuum state, and then the transition
probability is given by (\ref{transprob}) with $n=0$ and
$\varphi=\pi/\sqrt{2}$.

As another example, we consider again that at $t=0$ an atom is found in the
$\DN$ state, but let us now ask this question: If another atom comes between
$t$ and $t+\diff{t}$, with which probability will it be in the $\DN$ state as
well? Note the difference. Above we were interested in {\em the next\/} atom,
now in {\em another\/} atom. Clearly, the initial state is obtained by the
same reduction, viz.\ that of (\ref{reduction}), but now $\rho_t$ is
determined by the OAM master equation (\ref{diagME0}) with
$\varphi=\pi/\sqrt{2}$. Rather than (\ref{condoft}), we get here
\begin{equation}\label{anotherone}
\mbox{prob}_t(\UP\to\DN)=
\underbrace{\rule[-14pt]{0pt}{10pt}
\frac{A\sin^2(\pi/\sqrt{2})}{A+r\sin^2(\pi/\sqrt{2})}
}_{\mbox{\small a priori probability}}
\underbrace{\rule[-14pt]{0pt}{10pt}
\left(1-\exp\left[-\left(A+r\sin^2(\pi/\sqrt{2})\right)t\right]\right)
}_{\mbox{\small correlation function for $\DN$ atoms}}\,.
\end{equation}
We recognize that the first factor is the a priori probability of
(\ref{apriori}), and the second factor is the so-called correlation function.
It is a numerical measure for the strength of the correlations among atoms
emerging in the $\DN$ state. If the temporal separation is too large ---
physically: $t\gg1/A,1/r$ --- there must not be any correlations. Indeed, the
correlation function approaches unity for $t\to\infty$, and the conditional
probability (\ref{anotherone}) equals the a priori probability in this limit,
as it should.

These two examples must suffice as illustrations for the typical reasonings
that are necessary for the computation of statistical properties of the
emerging atoms. Let us just mention that the experimental data refers
actually to the statistics of the detector clicks, rather than to the atoms
themselves, and therefore the efficiencies of the detectors have to be taken
into account as well. Naturally, that introduces further complications which
are, however, under good theoretical control.

\SEC{Acknowledgments}

I would like to express my sincere gratitude towards the organizers and the
sponsors of the 19th Nathiagali Summer College for the very kind hospitality
I experienced in Pakistan and particularly in Nathiagali. I thank especially
M. S. Zubairy, K. A. Shoaib, and S. M. Farooqi for treating me so well. The
splendid company of I. Ashraf and S. Qamar is unforgettable.

\newpage

\tableofcontents

\listoffigures

\end{document}